\documentclass[12pt]{article}

\usepackage{amsfonts,amsmath}
\usepackage{graphicx}
\usepackage{appendix}
\usepackage[french,english]{babel}
\usepackage[utf8]{inputenc}

\renewcommand{\theequation}{\thesection.\arabic{equation}}

\newcommand\encadremath[1]{\vbox{\hrule\hbox{\vrule\kern8pt
\vbox{\kern8pt \hbox{$\displaystyle #1$}\kern8pt}
\kern8pt\vrule}\hrule}}
\def\enca#1{\vbox{\hrule\hbox{
\vrule\kern8pt\vbox{\kern8pt \hbox{$\displaystyle #1$}
\kern8pt} \kern8pt\vrule}\hrule}}

\newcommand\figureframex[3]{
\begin{figure}[bth]
\hrule\hbox{\vrule\kern8pt
\vbox{\kern8pt \vbox{
\begin{center}
{\mbox{\epsfxsize=#1.truecm\epsfbox{#2}}}
\end{center}
\caption{#3}
}\kern8pt}
\kern8pt\vrule}\hrule
\end{figure}
}
\newcommand\figureframey[3]{
\begin{figure}[bth]
\hrule\hbox{\vrule\kern8pt
\vbox{\kern8pt \vbox{
\begin{center}
{\mbox{\epsfysize=#1.truecm\epsfbox{#2}}}
\end{center}
\caption{#3}
}\kern8pt}
\kern8pt\vrule}\hrule
\end{figure}
}

\renewcommand{\thesection}{\arabic{section}}
\renewcommand{\theequation}{\arabic{section}-\arabic{equation}}
\makeatletter
\@addtoreset{equation}{section}
\makeatother
\newtheorem{theorem}{Theorem}[section]

\newtheorem{remark}{Remark}[section]
\newtheorem{proposition}{Proposition}[section]
\newtheorem{lemma}{Lemma}[section]
\newtheorem{corollary}{Corollary}[section]
\newtheorem{definition}{Definition}[section]
\def\br{\begin{remark}\rm\small}
\def\er{\end{remark}}
\def\bt{\begin{theorem}}
\def\et{\end{theorem}}
\def\bd{\begin{definition}}
\def\ed{\end{definition}}
\def\bp{\begin{proposition}}
\def\ep{\end{proposition}}
\def\bl{\begin{lemma}}
\def\el{\end{lemma}}
\def\bc{\begin{corollary}}
\def\ec{\end{corollary}}
\def\beaq{\begin{eqnarray}}
\def\eeaq{\end{eqnarray}}
\newcommand{\proof}[1]{{\noindent \bf proof:}\par
{#1} $\square$}

\newcommand{\td}{\tilde}

\newcommand{\beq}{\begin{equation}}
\newcommand{\eeq}{\end{equation}}
\newcommand{\beqq}{\begin{equation*}}
\newcommand{\eeqq}{\end{equation*}}
\newcommand{\bea}{\begin{eqnarray}}
\newcommand{\eea}{\end{eqnarray}}
\newcommand{\beaa}{\begin{eqnarray*}}
\newcommand{\eeaa}{\end{eqnarray*}}

\newcommand{\Tr}{\operatorname{Tr}}

\newcommand{\Res}{\mathop{\,\rm Res\,}}
\usepackage[pdftex]{hyperref}
\hypersetup{colorlinks,urlcolor=magenta,citecolor=red,linkcolor=blue,filecolor=black}

\begin{document}

\addtolength{\baselineskip}{0.20\baselineskip}
\begin{center}
\vspace{1cm}

{\Large \bf WKB solutions of difference equations and reconstruction by the topological recursion}

\vspace{1cm}

{Olivier Marchal}$^1$

\vspace{5mm}
$^1$\  Universit\'{e} de Lyon, CNRS UMR 5208, Universit\'{e} Jean Monnet,
\\
Institut Camille Jordan, France
\vspace{5mm} 

\end{center}


\abstract{The purpose of this article is to analyze the connection between Eynard-Orantin topological recursion and formal WKB solutions of a $\hbar$-difference equation: $\Psi(x+\hbar)=\left(e^{\hbar\frac{d}{dx}}\right) \Psi(x)=L(x;\hbar)\Psi(x)$ with $L(x;\hbar)\in GL_2( (\mathbb{C}(x))[\hbar])$. In particular, we extend the notion of determinantal formulas and topological type property proposed for formal WKB solutions of $\hbar$-differential systems to this setting. We apply our results to a specific $\hbar$-difference system associated to the quantum curve of the Gromov-Witten invariants of $\mathbb{P}^1$ for which we are able to prove that the correlation functions are reconstructed from the Eynard-Orantin differentials computed from the topological recursion applied to the spectral curve $y=\cosh^{-1}\frac{x}{2}$. Finally, identifying the large $x$ expansion of the correlation functions, proves a recent conjecture made by B. Dubrovin and D. Yang regarding a new generating series for Gromov-Witten invariants of $\mathbb{P}^1$.}


\tableofcontents

\section{Introduction}
\subsection{General context}
In the last decade, many interesting relations have been proved between integrable systems and Eynard-Orantin topological recursion. In particular, in a series of papers \cite{Deter,BBEnew,BE2}, the authors proved that starting from a linear differential system of the form:
\beq \label{general}\hbar \frac{d}{dx} \Psi(x;\hbar)= D(x;\hbar) \Psi(x;\hbar) \text{ with } D(x;\hbar)=\sum_{k=0}^\infty D_k(x)\hbar^k \in \mathcal{M}_d(\mathbb{C}(x))\footnote{$\mathcal{M}_d(\mathbb{C}(x))$ stands for the set of matrices of size $d\times d$ with entries in the field of rational functions of $x$} \eeq
where $\Psi(x;\hbar)$ are some formal WKB solutions, we may define, using appropriate determinantal formulas, correlation functions $W_n(x_1,\dots,x_n)$ that satisfy the same set of loop equations as those arising in the theory of the topological recursion and Hermitian random matrix models. Moreover, under some additional conditions on the differential system known as the topological type property, it was proved that these correlation functions identify with the corresponding Eynard-Orantin differentials $\left(\omega_n^{(g)}\right)_{g\geq 0, n\geq 1}$ computed from the application of the topological recursion to the classical spectral curve defined by $E(x,y)\overset{\text{def}}{=}\det(yI_d-D_0(x))=0$. We remind the reader that the Eynard-Orantion differentials $\omega_n^{(g)}(x_1,\dots,x_n)$ are defined in the following by the following recursion:
\begin{definition}[Topological recursion] Let $\Sigma$ be a Riemann surface of genus $g$ associated to the polynomial equation $E(x,y)=0$ and let $(x(z),y(z))$ with $z\in \Sigma$ be a parametrization of $\Sigma$. The branchpoints $(a_i)_{i\leq r}$ are defined as the zeros of the $1$-form $dx$. They are assumed to be simple zeros so that a local involution $z\mapsto \bar{z}$ such that $(x(z)=x(\bar{z}))$ is well-defined locally around each branchpoint. Let $\omega_2^{(0)}$ be the normalized bi-differential form (sometimes called Bergmann kernel) on $\Sigma$. The Eynard-Orantin differentials $\left(\omega_n^{(g)}\right)_{g\geq 0, n\geq 1}$ are defined recursively by:
\bea \omega_1^{(0)}&=&-ydx  \text{ and }\cr
\omega_n^{(g)}(z_1\dots,z_n)&=&\sum_{i=1}^r \Res_{z\to a_i} \frac{\mathcal{K}(z_1,z)}{\omega(q)}\Big( \omega_{n+2}^{(g-1)}(z,\bar{z},p_{I})\cr
&&+\sum_{m=0}^g \sum'_{I_1\sqcup I_2=I} \omega_{|I_1|+1}^{(m)}(z,z_{I_1}) \omega_{|I_2|+1}^{(g-m)}(\bar{z},z_{I_2}) \Big)
\eea 
where
\beq \omega(z)=(y(z)-y(\bar{z}))dx(z)\,\,\text{ and }\,\,\mathcal{K}(z_0,z)=\frac{1}{2}\int_z^{\bar{z}}\omega_2^{(0)}(s,z)\eeq
When $g=0$, the parametrization of $\Sigma$ may be chosen so that $z\in \bar{\mathbb{C}}=\mathbb{C}\cup\{\infty\}$ and $\omega_2^{(0)}$ is explicitly given by $\omega_2^{(0)}(z_1,z_2)=\frac{dz_1\otimes dz_2}{(z_1-z_2)^2}$.  
\end{definition}

 For a given differential system, proving the topological type property remains challenging, and only a few specific differential systems have been shown to satisfy it so far. For example, this strategy has been successfully used for Lax pairs associated to the six Painlev\'{e} equations \cite{P2,Painleve,IS,P5}. In particular, in the Painlev\'{e} context, the existence of an additional time differential system is crucial to prove the topological type property. Thus, the topological type property seems deeply related with some kind of integrability conditions. Recently, the general setup has been extended to general connections on Lie algebras \cite{LoopLie,LoopLie2} and the reconstruction of formal WKB solutions via the topological recursion has been proved for a wide class of rational matrices $D(x;\hbar)$ in \cite{Reconstruction}.\\
The connection between formal WKB expansion and Eynard-Orantin topological recursion is particularly interesting for the issue of ``quantizing'' the classical spectral curve. Indeed, under certain conditions, we expect that if we define the functions $\left(F_{g,n}\right)_{g\geq 0,n\geq 1}$:
\beq \label{Fgn}F_{g,n}(z_1,\dots,z_n)=\int^{z_1}\dots \int^{z_n} \omega_n^{(g)}(z_1',\dots,z_n')\eeq
then the formal wave function:
\beq \psi(x;\hbar)=\exp\left(\frac{1}{\hbar}F_{0,1}(x)+\frac{1}{2!}F_{0,2}(x,x)+\sum_{2g-2+n>0}\frac{\hbar^{2g-2+n}}{n!}F_{g,n}(x,\dots,x)\right)\eeq
satisfy a ``quantized'' version of the classical spectral curve:
\beq \label{QuantumGeneral}E\left(x,\hbar \frac{d}{dx}\right) \psi(x;\hbar)=0\eeq
Note that the quantization procedure requires to decide where to insert the operators $\hbar^k\frac{d^k}{dx^k}$ since they do not commute with functions of $x$, and also to choose the lower bounds in \eqref{Fgn} correctly (See \cite{QuantumP1} for the $\mathbb{P}^1$ case and \cite{Reconstruction} for general results). Details about relations between quantum curves and the topological recursion can be found in the review \cite{ReviewNorbury} or in \cite{Reconstruction} that covers a wide range of spectral curves. Since $E(x,y)$ is a monic polynomial of degree $d$ in the variable $y$, the corresponding quantum curve \eqref{QuantumGeneral} is an ordinary linear differential equation of degree $d$. Thus, $\psi(x;\hbar)$ may be seen as an entry of a $d\times d$ linear differential system of type \eqref{general} for which we expect that the topological type property holds. In other words, we have the following picture:

\bigskip
\bigskip

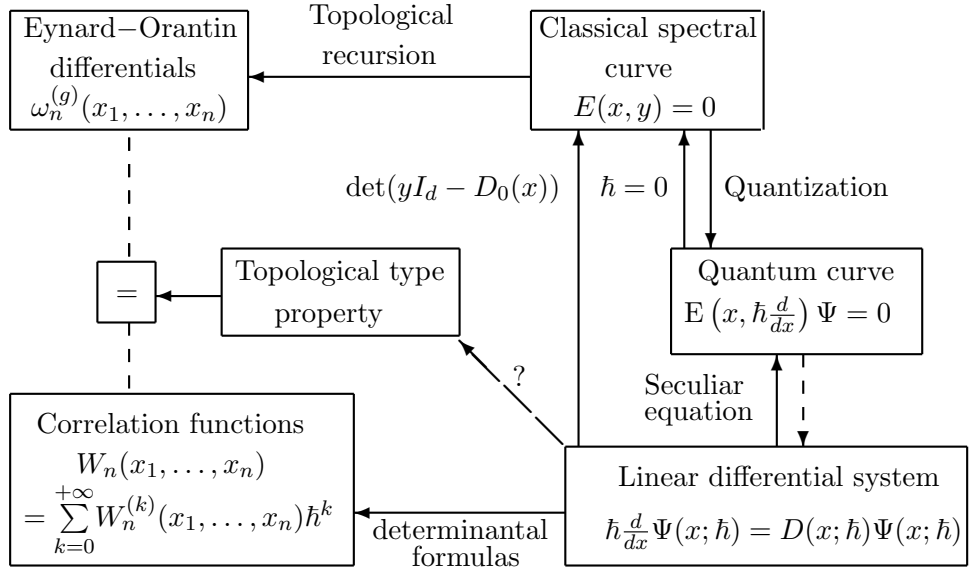
\begin{center}
\begin{picture}(365,210)
\thicklines 
\put(5,200){Eynard$-$Orantin} 
\put(15,185){differentials}
\put(8,170){$\omega_n^{(g)}(x_1,\dots,x_n)$}
\put(0,165){\line(1,0){90}}
\put(0,210){\line(1,0){90}}
\put(0,165){\line(0,1){45}}
\put(90,165){\line(0,1){45}}

\put(113,205){Topological} 
\put(118,190){recursion} 
\thicklines
\put(197,185){\vector(-1,0){107}}

\put(40,100){$=$}
\put(33,92){\line(1,0){22}}
\put(33,115){\line(1,0){22}}
\put(33,92){\line(0,1){23}}
\put(55,92){\line(0,1){23}}

\put(85,108){Topological type}
\put(100,92){property}
\put(80,87){\line(1,0){90}}
\put(80,120){\line(1,0){90}}
\put(80,87){\line(0,1){33}}
\put(170,87){\line(0,1){33}}
\put(80,102){\vector(-1,0){25}} 

\put(260,108){Quantum curve}
\put(255,92){$\text{E}\left(x,\hbar \frac{d}{dx}\right)\Psi=0$}
\put(250,80){\line(1,0){95}}
\put(250,120){\line(1,0){95}}
\put(250,80){\line(0,1){40}}
\put(345,80){\line(0,1){40}}

\put(270,140){Quantization}
\put(223,140){$\hbar=0$}
\put(255,120){\vector(0,1){45}}
\put(265,165){\vector(0,-1){45}}
\put(215,45){\vector(0,1){120}}

\put(127,140){$\text{det}(yI_d-D_0(x))$}

\put(200,200){Classical spectral} 
\put(225,185){curve}
\put(213,170){$E(x,y)=0$}
\put(197,165){\line(1,0){87}}
\put(197,210){\line(1,0){87}}
\put(197,165){\line(0,1){45}}
\put(285,165){\line(0,1){45}}

\put(230,30){Linear differential system} 
\put(225,10){$\hbar\frac{d}{dx}\Psi(x;\hbar)=D(x;\hbar)\Psi(x;\hbar)$}
\put(210,0){\line(1,0){152}}
\put(210,45){\line(1,0){152}}
\put(210,0){\line(0,1){45}}
\put(362,0){\line(0,1){45}}

\put(190,68){?}

\put(10,50){Correlation functions} 
\put(25,35){$W_n(x_1,\dots,x_n)$}
\put(5,15){$=\underset{k=0}{\overset{+\infty}{\sum}} W_n^{(k)}(x_1,\dots,x_n)\hbar^k$}
\put(0,0){\line(1,0){130}}
\put(0,65){\line(1,0){130}}
\put(0,0){\line(0,1){65}}
\put(130,0){\line(0,1){65}}

\put(137,10){\small{determinantal}}
\put(152,0){\small{formulas}}
\put(210,20){\vector(-1,0){80}} 

\put(240,65){Seculiar}
\put(240,55){equation}
\put(290,45){\vector(0,1){35}}

\multiput(45, 67)(0,10){3}{\line(0,1){5}}
\multiput(45, 117)(0,10){5}{\line(0,1){5}}
\multiput(45, 117)(0,10){5}{\line(0,1){5}}
\multiput(300,80)(0,-10){3}{\line(0,-1){5}}
\put(300,50){\vector(0,-1){5}}
\multiput(209,46)(-12,12){2}{\line(-1,1){10}}
\put(185,70){\vector(-1,1){15}}

\end{picture}
\end{center}
\bigskip

In the previous picture, the two dashed arrows represent the two steps for which the correspondence is not perfectly understood. For example, it is not known at the moment if the topological type property is a necessary condition or if it is only a sufficient condition to obtain the identification between the correlation functions and Eynard-Orantin differentials. Moreover, for a given linear scalar differential equation, there exist many compatible matrix differential systems and it is unclear if only one of them is compatible with the global picture presented above.\\
The main purpose of this article is to prove that a similar picture holds in the case of Gromov-Witten invariants of $\mathbb{P}^1$ for which the starting point is no longer a linear differential equation \eqref{QuantumGeneral} but rather a linear $\hbar$-difference equation:
\beq E\left(x,\text{exp}\left(\hbar\frac{d}{dx}\right)\right) \psi(x;\hbar)=0 \,\,,\,\, E(x,y) \text{ polynomial in } y\eeq
Consequently, we would like to extend the previous picture in the following way:
\bigskip
\begin{center}
\begin{picture}(400,310)
\thicklines 
\put(5,300){Eynard$-$Orantin} 
\put(15,285){differentials}
\put(8,270){$\omega_n^{(g)}(x_1,\dots,x_n)$}
\put(0,265){\line(1,0){90}}
\put(0,310){\line(1,0){90}}
\put(0,265){\line(0,1){45}}
\put(90,265){\line(0,1){45}}

\put(113,305){Topological} 
\put(118,290){recursion} 
\thicklines
\put(197,285){\vector(-1,0){107}}

\put(40,200){$=$}
\put(33,192){\line(1,0){22}}
\put(33,215){\line(1,0){22}}
\put(33,192){\line(0,1){23}}
\put(55,192){\line(0,1){23}}

\put(85,208){Topological type}
\put(100,192){property}
\put(80,187){\line(1,0){90}}
\put(80,220){\line(1,0){90}}
\put(80,187){\line(0,1){33}}
\put(170,187){\line(0,1){33}}
\put(80,202){\vector(-1,0){25}} 

\put(260,208){Quantum curve}
\put(255,192){$\text{E}\left(x,\text{exp}\left(\hbar \frac{d}{dx}\right)\right)\Psi=0$}
\put(250,180){\line(1,0){115}}
\put(250,220){\line(1,0){115}}
\put(250,180){\line(0,1){40}}
\put(365,180){\line(0,1){40}}

\put(270,240){Quantization}
\put(223,240){$\hbar=0$}
\put(255,220){\vector(0,1){45}}
\put(265,265){\vector(0,-1){45}}
\put(215,145){\vector(0,1){120}}

\put(127,240){$\text{det}(yI_d-D_0(x))$}

\put(200,300){Classical spectral} 
\put(225,285){curve}
\put(213,270){$E(x,e^{\,y})=0$}
\put(197,265){\line(1,0){87}}
\put(197,310){\line(1,0){87}}
\put(197,265){\line(0,1){45}}
\put(285,265){\line(0,1){45}}

\put(130,130){Linear differential system} 
\put(125,110){$\hbar\frac{d}{dx}\Psi(x;\hbar)=D(x;\hbar)\Psi(x;\hbar)$}
\put(110,100){\line(1,0){152}}
\put(110,145){\line(1,0){152}}
\put(110,100){\line(0,1){45}}
\put(262,100){\line(0,1){45}}

\put(130,165){?}

\put(10,50){Correlation functions} 
\put(25,35){$W_n(x_1,\dots,x_n)$}
\put(5,15){$=\underset{k=0}{\overset{+\infty}{\sum}} W_n^{(k)}(x_1,\dots,x_n)\hbar^k$}
\put(0,0){\line(1,0){130}}
\put(0,65){\line(1,0){130}}
\put(0,0){\line(0,1){65}}
\put(130,0){\line(0,1){65}}

\put(137,10){\small{determinantal}}
\put(152,0){\small{formulas}}
\put(210,20){\vector(-1,0){80}} 

\put(125,85){\small{determinantal}}
\put(135,75){\small{formulas}}
\put(120,100){\vector(0,-1){35}}

\put(200,70){exp}
\put(225,100){\vector(0,-1){55}}
\multiput(235, 45)(0,10){5}{\line(0,1){5}}
\put(235,95){\vector(0,1){5}}
\put(245,70){log}

\put(320,115){Seculiar}
\put(320,105){equation}
\put(310,45){\vector(0,1){135}}

\put(230,30){Linear $\hbar$-difference system} 
\put(215,10){$\left(\text{exp}\left(\hbar\frac{d}{dx}\right)\right)\Psi(x;\hbar)=L(x;\hbar)\Psi(x;\hbar)$}
\put(210,0){\line(1,0){183}}
\put(210,45){\line(1,0){183}}
\put(210,0){\line(0,1){45}}
\put(393,0){\line(0,1){45}}

\multiput(290,180)(0,-10){15}{\line(0,-1){5}}
\put(290,50){\vector(0,-1){5}}
\multiput(45, 67)(0,10){13}{\line(0,1){5}}
\multiput(45, 217)(0,10){5}{\line(0,1){5}}
\multiput(125, 148)(0,10){4}{\line(0,1){5}}

\end{picture}
\end{center}
\bigskip

More precisely, the paper focuses on the simpler case of a degree two $\hbar$-difference system of the form:
\beq \label{Setting}\delta_\hbar \Psi(x;\hbar)\overset{\text{def}}{=}\text{exp}\left(\hbar\frac{d}{dx}\right)\Psi(x;\hbar)=\sum_{k=0}^\infty \frac{\hbar^k}{k!}\frac{d^k}{dx^k}\Psi(x;\hbar)=L(x;\hbar)\Psi(x;\hbar)\eeq
where:
\beq L(x;\hbar)=\sum_{k=0}^\infty L_k(x)\hbar^k \in GL_2(\mathbb{C}(x))\eeq
In this article, we will use the notation ``$;\hbar$'' to stress that the quantity has to be understood as a formal WKB (or Taylor expansion is the term in $e^{\frac{1}{\hbar}}$ is vanishing) in $\hbar$. In this setting, the operator $\delta_\hbar$ (which does not obey Leibniz rule and thus is not a derivation operator) may be seen as a $\hbar$-difference operator:
\beq \delta_\hbar f(x;\hbar)=\sum_{k=0}^\infty \frac{\hbar^k}{k!}\frac{d^k}{dx^k}f(x;\hbar)=f(x+\hbar;\hbar)\eeq
Note in particular that the setting \eqref{Setting} may be seen as the exponentiated version of a standard linear differential system and therefore results presented in this article are likely to extend to general connections on some Lie groups $G$ (in the case of $\mathbb{P}^1$, the Lie group is $G=SL_2(\mathbb{C})$). Therefore, the article may also be seen as a first step to generalize results of \cite{LoopLie2} to problems directly defined on the Lie group rather than on the associated Lie algebra.

The main steps and results of the article are:
\begin{itemize} \item In section \ref{DiffDiff}, we start from the $\hbar$-difference equation satisfied by the wave function arising in the enumeration of Gromov-Witten invariants of $\mathbb{P}^1$ (proved in \cite{QuantumP1}) to build an adapted $2\times 2$ $\hbar$-difference system. Though this step may appear straightforward, the linear system has to be selected properly in order to satisfy the parity property of the topological type property.
\item In section \ref{Gene}, for a general $2\times 2$ $\hbar$-difference system $\delta_\hbar \Psi(x;\hbar)=L(x;\hbar)\Psi(x;\hbar)$, we provide the explicit relations with its associated compatible differential system $\hbar \frac{d}{dx} \Psi(x;\hbar)=D(x;\hbar)\Psi(x;\hbar)$. The connection is made in both ways, i.e. we explain how to derive $L(x;\hbar)$ from $D(x;\hbar)$ and vice versa how to obtain $D(x;\hbar)$ from $L(x;\hbar)$. The formulas are applied to the case of $\mathbb{P}^1$ in section \ref{AppliP11}.
\item In section \ref{MMSection}, we remind the definition of the matrix $M(x;\hbar)$ used in the definition of the determinantal formulas. In particular, we show how the matrix $M(x;\hbar)$ may be constructed directly from the matrix $L(x;\hbar)$. Explicit formulas are provided for the leading order $M_0(x)$ as well as a recursion to get all higher orders of the $\hbar$-expansion of $M(x;\hbar)$. We finally apply these formulas in the case of $\mathbb{P}^1$ and obtain a control over the singularities of the correlation functions in section \ref{AppliP11}.
\item In section \ref{DeterSection}, we remind the definitions of determinantal formulas (in the case of $2\times 2$ differential systems) and of the topological type property. We prove that the latter property is satisfied in the case of $\mathbb{P}^1$ in section \ref{ProofTT}.
\item In section \ref{Comparision}, we compare the matrix $M(x;\hbar)$ defined in section \ref{MMSection} with the conjecture proposed by Dubrovin and Yang in \cite{Dubrovin}. Both matrices are shown to be identical and thus we obtain a proof of their conjecture.  
\end{itemize}

\subsection{Known results about Gromov-Witten invariants of $\mathbb{P}^1$}
In this section, we briefly review the known results about Gromov-Witten invariants of $\mathbb{P}^1$ and present the conjecture of Dubrovin and Yang that we will prove in this article.

\subsubsection{Quantum curve associated to Gromov-Witten invariants of $\mathbb{P}^1$}
Let $\overline{\mathcal{M}}_{g,n}(\mathbb{P}^1,d)$ denote the moduli space of stable maps of degree $d$ from a $n$-pointed genus $g$ curve to $\mathbb{P}^1$ and define the descendant Gromov-Witten invariants of $\mathbb{P}^1$ by:
\beq \left< \underset{i=1}{\overset{n}{\prod}}\tau_{b_i}(\alpha_i)\right>^d_{g,n}=\int_{\left[\overline{\mathcal{M}}_{g,n}(\mathbb{P}^1,d)\right]^{\text{vir}} }\underset{i=1}{\overset{n}{\prod}} \psi_i^{\,b_i}ev_i^*(\alpha_i)\eeq
where $\left[\overline{\mathcal{M}}_{g,n}(\mathbb{P}^1,d)\right]^{\text{vir}}$ is the virtual fundamental class of the moduli space as defined in \cite{QuantumP1}. Using a formal parameter $\hbar$, we can regroup the Gromov-Witten invariants in a generating series:
\beq \left<\tau_{k_1}(\omega)\dots \tau_{k_n}(\omega)\right>=\sum_{2g-2+2d=k_1+\dots+k_n} \hbar^{2g-2+n}\left<\tau_{k_1}(\omega)\dots \tau_{k_n}(\omega)\right>_{g,n}^d\eeq
where $\omega\in H^2(\mathbb{P}^1,\mathbb{Q})$ is the Poincar\'{e} dual to the point class and $1$ is the generator of $H^0(\mathbb{P}^1,\mathbb{Q})$. In \cite{NS} the authors proved that genus-zero and genus-one stationary Gromov–Witten invariants of $\mathbb{P}^1$ arise as the Eynard–Orantin invariants of the spectral curve $y(x)=\text{cosh}^{-1}\left(\frac{x}{2}\right)$. This result has been refined to all orders in \cite{QuantumP1} where the main result (Theorem $1.1$) proves that if we define the following formal series expansions in $\frac{1}{x}$:
\bea S_0(x)&=&x-x\log x+\sum_{d=1}^\infty \left<-\frac{(2d-2)! \tau_{2d-2}(\omega)}{x^{2d-1}}\right>^d_{0,1}\cr
S_1(x)&=&-\frac{1}{2}\log x +\frac{1}{2}\sum_{d=0}^\infty \left<\left(-\frac{\tau_0(1)}{2}-\sum_{b=0}^\infty \frac{b!\tau_b(\omega)}{x^{b+1}}\right)^2\right>^d_{0,2}\cr
F_{g,n}(x_1,\dots,x_n)&=&\left<\underset{d=1}{\overset{n}{\prod}}\left(-\frac{\tau_0(1)}{2}-\sum_{b=0}^\infty \frac{b!\tau_b(\omega)}{x_i^{b+1}}\right)\right>^d_{g,n} \,,\text{ for } 2g-2+n>0\cr
&&
\eea
then the formal WKB wave function:
\beq \label{Wave}\psi(x;\hbar)=\exp\left(\frac{1}{\hbar}S_0(x)+S_1(x)+\sum_{2g-2+n>0}\frac{\hbar^{2g-2+n}}{n!}F_{g,n}(x,\dots,x)\right)\eeq
satisfies the $\hbar$-difference equation:
\beq \label{QuantumCurve} \left[\exp\left(\hbar \frac{d}{dx}\right)+\exp\left(-\hbar\frac{d}{dx}\right)-x\right]\psi(x,\hbar)=0\eeq
where the operator $\exp\left(\hbar \frac{d}{dx}\right)$ is defined by:
\beq \delta_\hbar\overset{\text{def}}{=}\exp\left(\hbar \frac{d}{dx}\right)\overset{\text{def}}{=}\sum_{k=0}^\infty \frac{\hbar^k}{k!}\frac{d^k}{dx^k} \eeq
Moreover, the functions $\left(F_{g,n}(x_1,\dots,x_n)\right)_{g\geq 0, n\geq 1, 2g+2-n>0}$ identify with the Eynard-Orantin differentials $\left(\omega_n^{(g)}\right)_{g\geq 0, n\geq 1, 2g+2-n>0}$ computed from the application of the topological recursion \cite{EO} on the genus $0$ classical spectral curve\footnote{The curve \eqref{ClassicalCurve} is of genus $0$ since it can be parametrized with a parameter $z$ living on the genus $0$ Riemann surface $\Sigma=\bar{\mathbb{C}}$. In particular, it has only $2$ branchpoints located at $z=\pm 1$ in the present parametrization}: 
\beq \label{ClassicalCurve} \forall\, z\in \bar{\mathbb{C}}\,:\, x(z)=z+\frac{1}{z} \,\,,\,\, y(z)=\ln(z) \,\,(\text{i.e. } y(x)=\text{cosh}^{-1}\left(\frac{x}{2}\right))\eeq
via the relations:
\beq F_{g,n}(z_1,\dots,z_n)=\int_0^{z_1}\dots \int_0^{z_n} \omega_n^{(g)}(z_1',\dots,z_n')\,\,,\,\,\forall\, g\geq 0, n\geq 1, 2g+2-n>0\eeq
In other words, the Eynard-Orantin differentials are related to the Gromov-Witten invariants of $\mathbb{P}^1$ by (see equation $2.9$ of \cite{QuantumP1}):
\beq \label{GWwng}\omega_n^{(g)}(x_1,\dots,x_n)=\sum_{k_1,\dots,k_n\geq 0} \frac{(k_1+1)!\dots (k_n+1)!}{x_1^{k_1+2}\dots x_n^{k_n+2}}\left<\tau_{k_1}(\omega)\dots \tau_{k_n}(\omega)\right>^d_{g,n}dx_1\dots dx_n\eeq

\begin{remark}
Note that the operator $\exp\left(\hbar \frac{d}{dx}\right)$ formally acts on functions via the Taylor formula:
\beq \exp\left(\hbar \frac{d}{dx}\right)f(x)=\sum_{k=0}^\infty \frac{f^{(k)}(x)}{k!}\hbar^k=f(x+\hbar)\eeq 
Hence the ``quantum curve'' \eqref{QuantumCurve} is often seen as a $\hbar$-difference equation:
\beq \label{DifferenceEquation}\psi(x+\hbar)+\psi(x-\hbar)-x\psi(x)=0\eeq
\end{remark}

\subsubsection{Dubrovin and Yang's conjecture\label{ConjectureSection}}

Recently, B. Dubrovin and D. Yang conjectured in \cite{Dubrovin} that the functions:
\bea
C_n(x_1,\dots,x_n;\hbar)&=&\sum_{k_1,\dots,k_n\geq 0} \frac{(k_1+1)!\dots (k_n+1)!}{x_1^{k_1+2}\dots x_n^{k_n+2}}\left<\tau_{k_1}(\omega)\dots \tau_{k_n}(\omega)\right>^d\cr
&=&\sum_{g=0}^\infty\hbar^{n-2-2g}\sum_{k_1,\dots,k_n\geq 0} \frac{(k_1+1)!\dots (k_n+1)!}{x_1^{k_1+2}\dots x_n^{k_n+2}}\left<\tau_{k_1}(\omega)\dots \tau_{k_n}(\omega)\right>_{g,n}^d\cr
&=& \sum_{g=0}^\infty\hbar^{n-2-2g}\frac{\omega_n^{(g)}(x_1,\dots,x_n)}{dx_1\dots dx_n}\cr
&&\text{(where $\omega_n^{(g)}$ are the Eynard-Orantin differentials associated }\cr
&&\text{ to the classical spectral curve \eqref{ClassicalCurve})}
\eea
may be formally reconstructed for $n\geq 2$ by:
\bea \label{C} C_2(x_1,x_2;\hbar)&=&\frac{\Tr\left(\td{M}(x_1;\hbar)\td{M}(x_2;\hbar)-1\right)}{(x_1-x_2)^2}\cr
C_n(x_1,\dots,x_n;\hbar)&=&\frac{(-1)^{n+1}}{n}\sum_{\sigma \in S_n}\frac{\Tr\left(\td{M}(x_{\sigma(1)};\hbar)\dots \td{M}(x_{\sigma(n)};\hbar)\right)}{(x_{\sigma(1)}-x_{\sigma(2)})\dots (x_{\sigma(n-1)}-x_{\sigma(n)})(x_{\sigma(n)}-x_{\sigma(1)})}\cr
\eea
where $\td{M}(x;\hbar)$ is a $2\times2$ matrix given by:
\beq \label{ConjectureM}\td{M}(x;\hbar)=\begin{pmatrix}1&0\\ 0&0\end{pmatrix}
+ \begin{pmatrix} \alpha(x;\hbar)&\beta(x;\hbar)\\ \gamma(x;\hbar)&-\alpha(x;\hbar)\end{pmatrix}
\eeq
with 
\small{\bea \label{ConjectureM2}\alpha(x;\hbar)&=&\sum_{j=0}^\infty \frac{1}{4^jx^{2j+2}}\sum_{i=0}^j \hbar^{2(j-i)}\frac{1}{i!(i+1)!}\sum_{l=0}^i(-1)^l(2i+1-2l)^{2j+1}\binom{2i+1}{l}\cr
\gamma(x;\hbar)&=&Q(x;\hbar)+P(x;\hbar)\cr
\beta(x;\hbar)&=&Q(x;\hbar)-P(x;\hbar)\cr
P(x;\hbar)&=&\sum_{j=0}^\infty \frac{1}{4^jx^{2j+1}}\sum_{i=0}^j \hbar^{2(j-i)}\frac{1}{(i!)^2}\sum_{l=0}^i (-1)^l (2i+1-2l)^{2j}\left[\binom{2i}{l}-\binom{2i}{l-1}\right]\cr
Q(x;\hbar)&=&\frac{1}{2}\sum_{j=0}^\infty \frac{1}{4^jx^{2j+2}}\sum_{i=0}^j\hbar^{2(j-i)+1}\frac{2i+1}{(i!)^2}\sum_{l=0}^i(-1)^l(2i+1-2l)^{2j}\left[\binom{2i}{l}-\binom{2i}{l-1}\right]\cr
&&
\eea}\normalsize{}

\begin{remark} Note that we have set the parameters $(\lambda,\epsilon)$ used in \cite{Dubrovin} to $(x,\hbar)$ in the previous formulas in order to match with the notations of \cite{QuantumP1} and \cite{Deter}. Moreover, we also followed the convention of \cite{QuantumP1} and included a power $\hbar^n$ in the definition of $\left<\tau_{k_1}(\omega)\dots \tau_{k_n}(\omega)\right>^d$. We also slightly modified the conjecture and added a factor $(-1)^n$ in the conjectured formula of $C_n(x_1,\dots,x_n)$ in order to match with the formalism of determinantal formulas of \cite{Deter}. Consequently, we also modified the sign of $Q(x;\hbar)$.
\end{remark}

We observe that the form of the conjecture is in perfect agreement with the determinantal formulas presented in \cite{Deter} (that were extended later in \cite{BBEnew} for $d\times d$ systems). Consequently, as explained above, the strategy of this article to prove the conjecture is the following:
\begin{enumerate}
\item Find a suitable $2\times 2$ differential systems $\hbar \frac{d}{dx} \Psi(x;\hbar)=D(x;\hbar)\Psi(x;\hbar)$ adapted to the problem.
\item Show that this system satisfies the topological type property. This implies that the functions $C_n(x_1,\dots,x_n)$ are indeed reconstructed by determinantal formulas similar to \eqref{C} but with a $2\times 2$ matrix $M(x;\hbar)$ defined from the differential system.
\item Show that the matrix $M(x;\hbar)$ is the same as the one proposed by Dubrovin and Yang. 
\end{enumerate}  

\section{WKB solutions of difference/differential systems}
In this section, we first show how to build a $2\times 2$ $\hbar$-difference system adapted to the formal WKB solution of the quantum curve \eqref{DifferenceEquation} arising in the enumeration of Gromov-Witten invariants of $\mathbb{P}^1$. Then, we propose to establish the connection between a general $2\times 2$ $\hbar$-difference system and its associated compatible $2\times 2$ $\hbar$-differential system. Finally, we apply the formulas to the $\mathbb{P}^1$ case.

\subsection{WKB solutions and matrix $L(x;\hbar)$: the case of $\mathbb{P}^1$\label{DiffDiff}}

Let us start from the quantum spectral curve \eqref{QuantumCurve}:
\beq \label{QuantumC}\left(\delta_\hbar+\delta_{-\hbar}-x\right)f(x)=0 \,\,\Leftrightarrow \,\, f(x+\hbar)+f(x-\hbar)-xf(x)=0 \eeq
where we have noted $\delta_\hbar$ the operator $\delta_\hbar=\underset{k=0}{\overset{\infty}{\sum}} \frac{\hbar^k}{k!}\frac{d^k}{dx^k}$. From the main theorem of \cite{QuantumP1}, we know that $\psi(x;\hbar)$ (given by equation \eqref{Wave}) is a formal WKB solution of equation \eqref{QuantumC}. Moreover, since the quantum curve \eqref{QuantumC} is invariant under the change $\hbar\to -\hbar$, we immediately get that 
\beq \label{phi} \phi(x;\hbar)=\exp\left(-\frac{1}{\hbar}S_0(x)+S_1(x)+\sum_{2g-2+n>0}(-1)^n\frac{\hbar^{2g-2+n}}{n!}F_{g,n}(x,\dots,x)\right)\eeq
is also another linearly independent WKB solution of \eqref{QuantumC}. Eventually, it is straightforward to observe that the matrix:
\beq \label{Psi}\Psi(x;\hbar)=\begin{pmatrix} \psi(x+\frac{\hbar}{2};\hbar)&\phi(x+\frac{\hbar}{2};\hbar)\\ \psi(x-\frac{\hbar}{2};\hbar)&\phi(x-\frac{\hbar}{2};\hbar)\end{pmatrix}\eeq
where $\psi(x\pm \frac{\hbar}{2})$ and $\phi(x\pm \frac{\hbar}{2})$ are to be understood as formal $\hbar$-WKB expansions:
\small{\bea \psi\left(x+\frac{\hbar}{2};\hbar\right)&=&\sum_{k=0}^\infty \frac{\hbar^k}{2^k k!}\frac{d^k}{dx^k}\psi(x;\hbar)=\exp\left(\frac{1}{\hbar}S_0(x)\right)\left(\sum_{k=1}^\infty \hat{S}_{k,+}(x)\hbar^{k-1}\right)\cr
\psi\left(x-\frac{\hbar}{2};\hbar\right)&=&\sum_{k=0}^\infty \frac{(-1)^k\hbar^k}{2^k k!}\frac{d^k}{dx^k}\psi(x;\hbar)=\exp\left(\frac{1}{\hbar}S_0(x)\right)\left(\sum_{k=1}^\infty \hat{S}_{k,-}(x)\hbar^{k-1}\right)\cr
\phi\left(x-\frac{\hbar}{2};\hbar\right)&=&\sum_{k=0}^\infty \frac{(-1)^k\hbar^k}{2^k k!}\frac{d^k}{dx^k}\phi(x;\hbar)=\exp\left(-\frac{1}{\hbar}S_0(x)\right)\left(\sum_{k=1}^\infty \td{S}_{k,-}(x)\hbar^{k-1}\right) \cr
\phi\left(x+\frac{\hbar}{2};\hbar\right)&=&\sum_{k=0}^\infty \frac{\hbar^k}{2^k k!}\frac{d^k}{dx^k}\phi(x;\hbar)=\exp\left(-\frac{1}{\hbar}S_0(x)\right)\left(\sum_{k=1}^\infty \td{S}_{k,+}(x)\hbar^{k-1}\right) \cr
&&
\eea}\normalsize{}
is a formal solution of the difference system:
\beq \label{DiffSystem} \delta_\hbar \Psi(x;\hbar)=\Psi(x+\hbar)=L(x;\hbar) \Psi(x;\hbar) \text{ with } L(x;\hbar)=\begin{pmatrix}x+\frac{\hbar}{2}&-1\\ 1&0\end{pmatrix}\eeq
Note in particular that the matrix $L(x;\hbar)$ does depend on the parameter $\hbar$ and that $\det L(x;\hbar)=1$. 

\begin{remark}\label{remarkdet} Since the space of solutions of \eqref{QuantumC} is a vector space of dimension $2$, one may choose any linear combinations of $\psi(x;\hbar)$ and $\phi(x;\hbar)$ as building blocks of the matrix $\Psi(x;\hbar)$. This is equivalent to perform a transformation of the form $\Psi(x;\hbar)\to \Psi(x;\hbar)C$ with a constant matrix $C\in GL_2(\mathbb{C}[\hbar])$. However, we chose in \eqref{Psi} very specific solutions of \eqref{QuantumC} since $\psi(x;\hbar)$ and $\phi(x;\hbar)$ involve only one of the exponential terms $\text{exp}\left(\pm \frac{S_0(x)}{\hbar}\right)$ whereas a generic linear combination would involve both in each entry of $\Psi(x;\hbar)$. Note that even if we impose the previous condition, the functions $\psi(x;\hbar)$ and $\phi(x;\hbar)$ are not uniquely determined since one could transform $(\psi(x;\hbar)),\phi(x;\hbar))$ to $(c_1(\psi(x;\hbar)),c_2\phi(x;\hbar))$ with $(c_1,c_2)\in \mathbb{C}[\hbar]^2$. This is equivalent to act on $\Psi(x;\hbar)$ by $\Psi(x;\hbar)\to \Psi(x;\hbar)\text{diag}(c_1,c_2)$. As we will see below, all interesting quantities ($D(x;\hbar)$, $M(x;\hbar)$, correlation functions, etc.) do not depend on the normalization of the matrix $\Psi(x;\hbar)$. In section \ref{SectionDeter}, we prove in the general setting that $\det \Psi(x;\hbar)$ is independent of $x$. Consequently, in the general setting, a natural choice is to normalize the determinant to $1$ using a transformation $(\psi(x;\hbar)),\phi(x;\hbar))$ to $(c_1(\psi(x;\hbar)),c_2\phi(x;\hbar))$ with $(c_1,c_2)\in \mathbb{C}[\hbar]^2$ even if this normalization is irrelevant for the results presented in this paper. In \eqref{Psi}, the normalization to $1$ is not imposed since we chose some specific entries, but it is clear with the previous discussion that it may be set to $1$ if desired.  
\end{remark}

\begin{remark} Another natural choice of combinations of wave functions $\psi(x;\hbar)$ and $\phi(x;\hbar)$ may have been:
\beqq \Psi_0(x;\hbar)=\begin{pmatrix} \psi(x;\hbar)&\phi(x;\hbar)\\ \psi(x-\hbar;\hbar)&\phi(x-\hbar;\hbar)\end{pmatrix}\eeqq
or more generally for any $\lambda\in \mathbb{C}$:
\beqq \Psi_\lambda(x;\hbar)=\begin{pmatrix} \psi(x+\lambda \hbar;\hbar)&\phi(x+\lambda \hbar;\hbar)\\ \psi(x+(\lambda-1) \hbar;\hbar)&\phi(x+(\lambda-1) \hbar;\hbar)\end{pmatrix}\eeqq
In particular, this choice is equivalent to a $\hbar$-difference system given by: 
\beqq L_\lambda(x;\hbar)=\begin{pmatrix} x+\lambda \hbar & -1\\ 1&0\end{pmatrix}\eeqq
However, as we will see in section \ref{SectionParity}, only $\lambda=\frac{1}{2}$ provides a system satisfying the parity property required for the topological type property. In other words, it is the only choice for which the determinantal formulas give rise to correlation functions $W_n(x_1,\dots,x_n)$ that may only involve even (resp. odd) powers of $\hbar$ when $n$ is even (resp. odd). Since this property is absolutely necessary to match with the Eynard-Orantin differentials (that always satisfy this property), then the only choice for $\Psi(x;\hbar)$ is precisely \eqref{Psi}. 
\end{remark}

\subsection{Generalization to arbitrary $2\times 2$ difference systems \label{Gene}}
The previous situation and most of the forthcoming results may be extended to some general $2\times 2$ $\hbar$-difference systems. Indeed, we first observe that the $\hbar$-difference system:
\beq \label{GeneralSe} \delta_\hbar \Psi(x;\hbar)=L(x;\hbar)\Psi(x;\hbar)\eeq
with $L(x;\hbar)$ a formal series in $\hbar$ is equivalent to say that the entries of $\Psi(x;\hbar)$ satisfy the seculiar equations:
\bea 0&=&L_{1,2}(x-\hbar)\Psi_{1,1}(x+\hbar;\hbar)+L_{1,2}(x;\hbar)(\det L(x-\hbar;\hbar))\Psi_{1,1}(x-\hbar;\hbar)\cr
&&-\left(L_{1,1}(x;\hbar)L_{1,2}(x-\hbar;\hbar)+L_{1,2}(x;\hbar)L_{2,2}(x-\hbar;\hbar)\right)\Psi_{1,1}(x;\hbar)\cr  
0&=&L_{1,2}(x-\hbar)\Psi_{1,2}(x+\hbar;\hbar)+L_{1,2}(x;\hbar)(\det L(x-\hbar;\hbar))\Psi_{1,2}(x-\hbar;\hbar)\cr
&&-\left(L_{1,1}(x;\hbar)L_{1,2}(x-\hbar;\hbar)+L_{1,2}(x;\hbar)L_{2,2}(x-\hbar;\hbar)\right)\Psi_{1,2}(x;\hbar)\cr 
\Psi_{2,1}(x;\hbar)&=& \frac{1}{L_{1,2}(x;\hbar)}\left[\Psi_{1,1}(x+\hbar;\hbar)-L_{1,1}(x;\hbar)\Psi_{1,1}(x;\hbar)\right]\cr
\Psi_{2,2}(x;\hbar)&=& \frac{1}{L_{1,2}(x;\hbar)}\left[\Psi_{1,2}(x+\hbar;\hbar)-L_{1,1}(x;\hbar)\Psi_{1,2}(x;\hbar)\right]
\eea
Therefore, if we define two independent WKB solutions $(\psi(x;\hbar),\phi(x;\hbar))$ of the $\hbar$-difference equation:
\bea 0&=& \Big[L_{1,2}(x-\hbar;\hbar)\delta_{\hbar}+L_{1,2}(x;\hbar)(\det L(x-\hbar;\hbar))\delta_{-\hbar}\cr
&&-L_{1,1}(x;\hbar)L_{1,2}(x-\hbar;\hbar)-L_{1,2}(x;\hbar)L_{2,2}(x-\hbar;\hbar)\Big]y=0\eea
with:
\bea \psi(x;\hbar)&=&\text{exp}\left(\frac{1}{\hbar}R_{-1}(x)+\sum_{k=0}^\infty R_k(x)\hbar^k\right)\cr
\phi(x;\hbar)&=&\text{exp}\left(-\frac{1}{\hbar}R_{-1}(x)+\sum_{k=0}^\infty S_k(x)\hbar^k\right)
\eea
then, for any $\lambda\in \mathbb{C}$, the matrix:
\footnotesize{\beq \Psi_\lambda(x;\hbar)=\begin{pmatrix} \psi(x+\lambda\hbar;\hbar)& \phi(x+\lambda\hbar;\hbar)\\ \frac{\psi(x+(\lambda+1)\hbar;\hbar)-L_{1,1}(x+\lambda\hbar;\hbar)\psi(x+\lambda\hbar;\hbar)}{L_{1,2}(x+\lambda\hbar;\hbar)}&  \frac{\phi(x+(\lambda+1)\hbar;\hbar)-L_{1,1}(x+\lambda\hbar;\hbar)\phi(x+\lambda\hbar;\hbar)}{L_{1,2}(x+\lambda\hbar;\hbar)}\end{pmatrix} \eeq}\normalsize{}
satisfies:
\beq \label{GeneralSelambda} \delta_\hbar \Psi_\lambda(x;\hbar)=L(x+\lambda \hbar;\hbar)\Psi_\lambda(x;\hbar)\overset{\text{def}}{=}L_\lambda(x;\hbar)\Psi_\lambda(x;\hbar)\eeq
Note that so far the choice of $\lambda\in \mathbb{C}$ is arbitrary. However, as in the $\mathbb{P}^1$ case, $\lambda$ may be used to impose some extra conditions on $\Psi_\lambda(x;\hbar)$, for example regarding the change $\hbar\leftrightarrow -\hbar$.

\begin{definition}[$\hbar$-difference system associated to a matrix $L(x;\hbar)$]\label{GeneralLSys} Let $L(x;\hbar)\in \mathcal{M}_d(\mathbb{C}(x)[\hbar])$ be a given matrix with $\det L(x;\hbar)=1$ and satisfying $L_{1,2}(x;\hbar=0)\neq 0$. Let $\psi(x;\hbar)$ and $\phi(x;\hbar)$ be two linearly independent formal WKB solutions of the $\hbar$-difference equation\footnote{A straightforward computation shows that such formal solutions exist since one may determine recursively the coefficients of their formal $\hbar$-expansion $\psi(x;\hbar)=e^{\frac{1}{\hbar} \psi_{-1}(x)}\left(\psi_0(x)+\underset{k=1}{\overset{\infty}{\sum}}\psi_k(x)\hbar^k\right)$. However, these solutions are only formal in the sense that the convergence of the series in some non-trivial domain is not guaranteed}:
\beaa 0&=&\Big[L_{1,2}(x-\hbar;\hbar)\delta_{\hbar}+L_{1,2}(x;\hbar)\delta_{-\hbar}\cr
&&+L_{1,1}(x;\hbar)L_{1,2}(x-\hbar;\hbar)+L_{1,2}(x;\hbar)L_{2,2}(x-\hbar;\hbar)\Big]y
\eeaa
Then, for any $\lambda\in \mathbb{C}$, we define the formal wave matrix:
\footnotesize{\beq \Psi_\lambda(x;\hbar)=\begin{pmatrix} \psi(x+\lambda\hbar;\hbar)& \phi(x+\lambda\hbar;\hbar)\\ \frac{\psi(x+(\lambda+1)\hbar;\hbar)-L_{1,1}(x+\lambda\hbar;\hbar)\psi(x+\lambda\hbar;\hbar)}{L_{1,2}(x+\lambda\hbar;\hbar)}&  \frac{\phi(x+(\lambda+1)\hbar;\hbar)-L_{1,1}(x+\lambda\hbar;\hbar)\phi(x+\lambda\hbar;\hbar)}{L_{1,2}(x+\lambda\hbar;\hbar)}\end{pmatrix} \eeq}\normalsize{}
that formally satisfies the $2\times 2$ $\hbar$-difference system:
\beqq \delta_\hbar \Psi_\lambda(x;\hbar)=L(x+\lambda\hbar ;\hbar)\Psi_\lambda(x;\hbar)\overset{\text{def}}{=}L_\lambda(x;\hbar)\Psi_\lambda(x;\hbar)\eeqq  
\end{definition}
 
Though the article focuses on the case of the specific system \eqref{DiffSystem}, results of sections \ref{SectionDeter} and \ref{DL}, are sufficiently general to apply for any system defined in \ref{GeneralLSys}. Eventually, we note that the $\hbar$-difference system \eqref{GeneralSe} may be seen as a special case of a linear differential system $\nabla_\hbar \Psi=0$ satisfied by a flat section $\Psi$ in a principal $G$-bundle over a complex curve equipped with a connection $\nabla_\hbar$. In our setting the Lie group $G$ is $SL_2(\mathbb{C})$ (since $\det L(x;\hbar)=1$) and the connection $\nabla_\hbar$ is the operator $\delta_\hbar=\text{exp}\left(\hbar \frac{d}{d\hbar}\right)$. In particular, in contrast with \cite{LoopLie}, the operator $\delta_\hbar=\text{exp}(\hbar \frac{d}{dx})$ directly comes in its exponentiated form while in \cite{LoopLie}, the starting point was the operator $\hbar \frac{d}{dx}$ defined on the corresponding Lie algebra $\mathfrak{g}$ (in our case $\mathfrak{g}=\mathfrak{s}l_2(\mathbb{C})$). In fact, a by-product of this article is also to show, on a simple example, that the reconstruction of the determinantal formulas via the topological recursion developed in \cite{LoopLie,P2,Painleve} in the Lie algebra setting may be adapted to similar problems defined directly on a Lie group.

\subsection{Properties of $\det \Psi(x;\hbar)$ \label{SectionDeter}}
 
Let us first observe that at the algebraic level, the operator $\delta_\hbar$ does not act as a derivation operator. Indeed, it satisfies the following algebraic rules: for any analytic function $f$ and $g$ and any complex number $\gamma$:
\beaa \delta_\hbar(\gamma)&=&\gamma \cr
\delta_\hbar(fg)(x)&=&\delta_\hbar(f)(x)\delta_\hbar(g)(x) \,\,\, (\Leftrightarrow\,\,  (fg)(x+\hbar)=f(x+\hbar)g(x+\hbar)) \cr
\delta_\hbar(\gamma f+g)(x)&=&\gamma\delta_\hbar(f)(x)+\delta_\hbar(g)(x)\cr
&&  \,\,\, (\Leftrightarrow\,\,  (\gamma f+g)(x+\hbar)=\gamma f(x+\hbar)+g(x+\hbar))
\eeaa
Therefore, let us denote $s(x;\hbar)=\det \Psi(x;\hbar)$ where by definition, $s(x;\hbar)$ has to be understood as a formal WKB expansion in $\hbar$. However, from the definition of the matrix $\Psi(x;\hbar)=\Psi_{\lambda=0}(x;\hbar)$, it is obvious that the determinant does not contain terms proportional to $e^{\frac{1}{\hbar}}$. Thus, it is simply a formal Taylor expansion in $\hbar$:
\beq s(x;\hbar)=\sum_{k=0}^\infty s_k(x)\hbar^k\eeq  
From the fact that $\det L(x;\hbar)=1$, we obtain that:
\beq \label{sss} s(x+\hbar;\hbar)=s(x;\hbar) \,\, \Leftrightarrow \sum_{j=1}^\infty\sum_{i=0}^\infty \frac{1}{j!}\frac{d^j}{dx^j}s_i(x)\hbar^{i+j}=0\eeq

In particular, identifying the coefficient of order $\hbar^1$ in \eqref{sss} implies that $s_0'(x)=0$, i.e. that $s_0(x)$ does not depend on $x$. For $k\geq 1$, identifying the coefficient in $\hbar^{k+1}$ in \eqref{sss} shows that $s'_k(x)$ equals a linear combination of derivatives of the $(s_i)_{i\leq k-1}$. Therefore a trivial induction shows that:
\beq \forall \,k\geq 0\,:\, s_k'(x)=0\eeq
Consequently we obtain that $\det \Psi(x;\hbar)$ does not depend on $x$. Since $\psi(x;\hbar)$ and $\phi(x;\hbar)$ are determined up to a multiplicative constant (Cf. Remark \ref{remarkdet}), we may normalize them so that $\det \Psi(x;\hbar)=1$.

\begin{proposition}\label{PropDet} We may choose the solutions $\psi(x;\hbar)$ and $\phi(x;\hbar)$ so that $\det \Psi(x;\hbar)=1$. In particular, the determinant does not depend on $x$.  
\end{proposition}

From now on, we will always assume that the solutions $\psi(x;\hbar)$ and $\phi(x;\hbar)$ have been chosen so that $\det \Psi(x;\hbar)=1$.

\begin{remark}Note that the determinant condition and the condition of having only one diverging (when $\hbar\to 0$) exponential term in each entry of $\Psi(x;\hbar)$ do not uniquely determine the matrix $\Psi(x;\hbar)$ yet (or equivalently the solutions $\psi(x;\hbar)$ and $\phi(x;\hbar)$). Indeed, normalizing the determinant uses only one degree of freedom in the transformations $\Psi(x;\hbar)\to \Psi(x;\hbar)\text{diag}(c_1,c_2)$. Therefore, one could impose an additional normalizing condition on $\Psi(x;\hbar)$ (or equivalently on $\psi(x;\hbar)$ or $\phi(x;\hbar)$) in order to obtain uniqueness of the $\Psi(x;\hbar)$ matrix. For example, we may impose that $\psi(x;\hbar)$ take a specific value at a given point $x_0\in \mathbb{C}$. However, as mentioned earlier these normalization issues are irrelevant for our purpose since all interesting quantities ($D(x;\hbar)$, $M(x;\hbar)$, correlation functions, etc.) defined below do not depend on these normalizations. Hence for simplicity, we choose not to impose any additional artificial normalizing condition. 
\end{remark}

\subsection{The corresponding differential operator $D(x;\hbar)$\label{DL}}

As explained earlier, the main difference with the setting of \cite{LoopLie,P2,Painleve} is that the $\hbar$-difference system \eqref{DiffSystem} is defined on the Lie group $SL_2(\mathbb{C})$ with an exponential operator $\delta_\hbar=\text{exp}(\hbar \frac{d}{dx})$ rather than on the corresponding Lie algebra $\mathfrak{s}l_2(\mathbb{C})$. At the level of representation matrices $L(x;\hbar)$, we would like to find a $2\times 2$ matrix $D(x;\hbar)\in \mathfrak{s}l_2(\mathbb{C})$ such that:
\beq \label{Compat} \hbar \frac{d}{dx} \Psi(x;\hbar)=D(x;\hbar)\Psi(x;\hbar) \,\, \Leftrightarrow \,\, \delta_\hbar \Psi(x;\hbar)=\Psi(x+\hbar)=L(x;\hbar)\Psi(x;\hbar)\eeq
with $D(x;\hbar)=\underset{k=0}{\overset{\infty}{\sum}}D_k(x)\hbar^k$ (i.e. a formal series expansion in $\hbar$). Then, we observe that by definition:
\beq D(x;\hbar)=\left(\hbar\frac{d}{dx} \Psi(x;\hbar)\right)\Psi(x;\hbar)^{-1}\eeq
so that:
\beq \label{TRDD}\Tr D(x;\hbar)=\hbar \frac{1}{\det \Psi(x;\hbar)}\frac{d}{dx}(\det \Psi(x;\hbar))=\hbar\frac{d}{dx}\left(\ln\det\Psi(x;\hbar)\right)=0\eeq
since $\det(\Psi(x;\hbar))$ does not depend on $x$ (Proposition \ref{PropDet}). 
The compatibility of the two systems \eqref{Compat} and the fact that $\left[\delta_\hbar,\hbar \frac{d}{dx}\right]=0$ is equivalent to say that:
\bea\label{ConnectionLD} 0&=&\hbar L'(x;\hbar)+L(x;\hbar)D(x;\hbar)-(\delta_\hbar D(x;\hbar))L(x;\hbar)\cr
\delta_\hbar D(x;\hbar)&=&\sum_{i=0}^\infty\left(\sum_{j=0}^i \frac{\frac{d^j}{dx^j}D_{(i-j)}(x)}{j!}\right)\hbar^i
\eea

\subsubsection{General reconstruction of $L(x;\hbar)$ from $D(x;\hbar)$}
Reconstructing $L(x;\hbar)$ from the knowledge of $D(x;\hbar)$ can be done by induction and the knowledge that $L_0(x)=\text{exp}(D_0(x))$. Indeed we have:
\beq L(x;\hbar)\Psi(x;\hbar)=\sum_{k=0}^\infty \frac{\hbar^k}{k!}\frac{d^k}{dx^k}\Psi(x;\hbar)\eeq
Let us define the sequence of functions $(A_{k}(x;\hbar))_{k\geq 0}$ by:
\bea A_0(x;\hbar)&=&1\cr
A_1(x;\hbar)&=&D(x;\hbar)\cr
A_k(x;\hbar)&=&\sum_{i=0}^{k-1}\binom{k-1}{i} \left(\frac{d^{k-1-i}}{dx^{k-1-i}}D(x;\hbar)\right)A_i(x;\hbar)\hbar^{k-1-i}\,\,,\,\,\forall\, k\geq 2\cr
&&
\eea
Then, from the chain rule, we have for all $k\geq 0$:
\beq \hbar^k\frac{d^k}{dx^k}\Psi(x;\hbar)=A_k(x;\hbar) \Psi(x;\hbar)\eeq
Thus, we find:
\beq \label{AAAA} L(x;\hbar)=\sum_{k=0}^\infty \frac{1}{k!}A_k(x;\hbar) \eeq
Note that the functions $\left(A_k(x;\hbar)\right)_{k\geq 0}$ are (non-commutative) polynomials of $\left(D_j(x)\right)_{j\leq k}$ and their derivatives.

\subsubsection{General reconstruction of $D(x;\hbar)$ from $L(x;\hbar)$\label{blah}}
In this section, we propose to show how we can reconstruct the matrix $D(x;\hbar)$ from the knowledge of the matrix $L(x;\hbar)$ introduced in the general setting of Definition \ref{GeneralLSys}. In particular, we show for $2\times 2$ systems how to obtain the orders $D_k(x)$ with $k\geq 1$ by induction using \eqref{ConnectionLD}. The construction being completely algebraic, the following results may be used also for the shifted version $D_\lambda(x)$ and $L_\lambda(x)$. Since $\Tr D(x)=0$, we only need to determine entries $\left(D_k(x)\right)_{1,1}, \left(D_k(x)\right)_{1,2}$ and $\left(D_k(x)\right)_{2,1}$ for all $k\geq 1$. Projecting \eqref{ConnectionLD} at order $\hbar^k$ provides the following equation:
\bea \left[L_0(x),D_k(x)\right]&=&\sum_{j=1}^k \frac{1}{j!} D_{k-j}(x)+\sum_{l=1}^k\sum_{j=0}^{k-l} \frac{1}{j!}D^{(j)}_{k-l-j}(x)L_l(x)\cr
&&-L_{k-1}'(x)-\sum_{j=0}^{k-1}L_{k-j}(x)D_j(x) \eea
For clarity, we introduce the following notation:
\beq \text{O}_k(x)\overset{\text{def}}{=}\sum_{j=1}^k \frac{1}{j!} D_{k-j}(x)+\sum_{l=1}^k\sum_{j=0}^{k-l} \frac{1}{j!}D^{(j)}_{k-l-j}(x)L_l(x)-L_{k-1}'(x)-\sum_{j=0}^{k-1}L_{k-j}(x)D_j(x)\eeq
In particular, we observe that $\text{O}_k(x)$ only involves lower orders $\left(D_i(x)\right)_{i\leq k-1}$ of the $\hbar$-expansion of the $D(x)$ matrix. Therefore, to obtain an induction process, we need to find a way to extract $D_k(x)$ from the commutator $\left[L_0(x),D_k(x)\right]$. We find for the entries:
\bea \label{ThreeTwoRelations} (L_0(x))_{1,2}(D_k(x))_{2,1}-(L_0(x))_{2,1}(D_k(x))_{1,2} &=& (\text{O}_k(x))_{1,1}\cr
\left((L_0(x))_{1,1}-(L_0(x))_{2,2}\right) (D_k(x))_{1,2}-2(L_0(x))_{1,2}(D_k(x))_{1,1} &=& (\text{O}_k(x))_{1,2}\cr
\left((L_0(x))_{1,1}-(L_0(x))_{2,2}\right) (D_k(x))_{2,1}-2(L_0(x))_{2,1}(D_k(x))_{1,1} &=& -(\text{O}_k(x))_{2,1}\cr
&&
\eea
Unfortunately, only two of the former three equations are linearly independent so that we may only determine two of the three entries. Consequently, we need to find a third independent equation. This may be done using the projection of \eqref{ConnectionLD} at order $\hbar^{k+1}$. Indeed, it provides:
\bea \label{OrderHkPlus1} \left[L_0(x),D_{k+1}(x)\right]&=&D_k'(x)L_0(x)+\left[D_k(x),L_1(x)\right]-L_k'(x)+\sum_{j=2}^k \frac{1}{j!}D_{k+1-j}^{(j)}(x)L_0(x)\cr
&&-\sum_{j=0}^{k-1}L_{k+1-j}(x)D_j(x)+\sum_{l=1}^{k+1}\sum_{j=0\,, \,(l,j)\neq(1,0)}^{k+1-l}\frac{1}{j!}D^{(j)}_{k+1-l-j}(x)L_l(x)\cr
&&
\eea
For clarity, let us denote:
\bea R_{k-1}(x)&\overset{\text{def}}{=}&L_k'(x)-\sum_{j=2}^k \frac{1}{j!}D_{k+1-j}^{(j)}(x)L_0(x)+\sum_{j=0}^{k-1}L_{k+1-j}(x)D_j(x)\cr
&&-\sum_{l=1}^{k+1}\sum_{j=0\,, \,(l,j)\neq(1,0)}^{k+1-l}\frac{1}{j!}D^{(j)}_{k+1-l-j}(x)L_l(x)
\eea
so that \eqref{OrderHkPlus1} is equivalent to:
\beq \label{OrderHkPlus2} \left[L_0(x),D_{k+1}(x)\right]=D_k'(x)L_0(x)+\left[D_k(x),L_1(x)\right]-R_{k-1}(x)\eeq
Note also that $R_{k-1}(x)$ only involves orders $\left(D_i(x)\right)_{i\leq k-1}$. Since the trace of a commutator always vanishes, taking the trace of equation \eqref{OrderHkPlus2} gives:
\beq \Tr\left(D_k'(x)L_0(x)\right)=\Tr\left(R_{k-1}(x)\right)\eeq
In terms of entries of the matrices $D_k(x)$, this is equivalent to:
\bea \label{ThirdEq}\Tr R_{k-1}(x)&=&\left((L_0(x))_{1,1}-(L_0(x))_{2,2} \right) (D_k'(x))_{1,1}\cr
&&+(L_0(x))_{2,1} (D_k'(x))_{1,2}+(L_0(x))_{1,2} (D_k'(x))_{2,1}\eea
Therefore, combining \eqref{ThreeTwoRelations} and \eqref{ThirdEq} we obtain the system of equations:
\bea \label{3Eq}(\text{O}_k(x))_{1,1}&=&(L_0(x))_{1,2}(D_k(x))_{2,1}-(L_0(x))_{2,1}(D_k(x))_{1,2}\cr
(\text{O}_k(x))_{1,2}&=&\left((L_0(x))_{1,1}-(L_0(x))_{2,2}\right) (D_k(x))_{1,2}-2(L_0(x))_{1,2}(D_k(x))_{1,1}\cr
\Tr R_{k-1}(x)&=&\left((L_0(x))_{1,1}-(L_0(x))_{2,2} \right) (D_k'(x))_{1,1}+(L_0(x))_{2,1} (D_k'(x))_{1,2}\cr
&&+(L_0(x))_{1,2} (D_k'(x))_{2,1}\cr
&&
\eea
Using the fact that $(L_0)_{1,2}\neq 0$, we may extract $(D_k(x))_{1,1}$ and $(D_k(x))_{2,1}$ in terms of $(D_k(x))_{1,2}$ using the first two equations:
\bea \label{OtherEntries}(D_k(x))_{2,1}&=&\frac{(L_0(x))_{2,1}}{(L_0(x))_{1,2}}(D_k(x))_{1,2}+ \frac{(O_k(x))_{1,1}}{(L_0(x))_{1,2}}\cr
 (D_k(x))_{1,1}&=&\frac{(L_0(x))_{1,1}-(L_0(x))_{2,2}}{2(L_0(x))_{1,2}}(D_k(x))_{1,2}-\frac{(O_k(x))_{1,2}}{2(L_0(x))_{1,2}}\cr
&&\eea
Inserting these relations in the third equation of \eqref{3Eq} gives the following linear ODE for $(D_k(x))_{1,2}$:
\bea\label{ODEDk} &&\left[\left((L_0(x))_{1,1}-(L_0(x))_{2,2}\right)^2+4(L_0(x))_{1,2}(L_0(x))_{2,1}\right] (D_k'(x))_{1,2}\cr
&&+\Big[2(L_0(x))_{1,2}\left(\frac{(L_0(x))_{1,1}-(L_0(x))_{2,2}}{2(L_0(x))_{1,2}}\right)'\left((L_0(x))_{1,1}-(L_0(x))_{2,2}\right)\cr
&&+2\left((L_0(x))_{1,2}\right)^2\left(\frac{(L_0(x))_{2,1}}{(L_0(x))_{1,2}}\right)'\Big](D_k(x))_{1,2}\cr
&&=-2(L_0(x))_{1,2}\left((L_0(x))_{1,1}-(L_0(x))_{2,2}\right)\left(\frac{(O_k(x))_{1,2}}{2(L_0(x))_{1,2}}\right)' \cr
&&+2\left((L_0(x))_{1,2}\right)^2\left(\frac{(O_k(x))_{1,1}}{(L_0(x))_{1,2}}\right)'\cr
&&
\eea
We now observe that since $\det L_0(x)=1$ (because $\det L(x;\hbar)=1$), the coefficient in front of $(D_k'(x))_{1,2}$ is precisely the discriminant of the classical spectral curve:
\bea \left((L_0(x))_{1,1}-(L_0(x))_{2,2}\right)^2+4(L_0(x))_{1,2}(L_0(x))_{2,1} &=&\left(\Tr L_0(x)\right)^2-4\det L_0(x)\cr
&=& \left(\Tr L_0(x)\right)^2-4\eea
Therefore, using standard results on linear ODEs, we obtain the following proposition:

\begin{proposition}\label{PropSing} Equations \eqref{OtherEntries} and \eqref{ODEDk} as well as $\Tr D_k(x)=0$ determine recursively the matrices $\left(D_{j}(x)\right)_{j\geq 1}$ from the knowledge of $D_0(x)$ and $\left(L_i(x)\right)_{i\leq k}$. In particular, $x\mapsto D_k(x)$ may only have singularities at points corresponding to either singularities of the entries of the matrices $\left(L_i(x)\right)_{i\geq 0}$ or singularities of the entries of the matrix $D_0(x)$ or zeros of $(L_0(x))_{1,2}$. 
\end{proposition}

\subsubsection{General connection between $L_0(x)$ and $D_0(x)$}
Using the results of the previous sections and in particular \eqref{AAAA}, we find that at order $\hbar^0$, we have the relation:
\beq \label{Expo}\sum_{k=0}^\infty \frac{\frac{d^k}{dx^k}D_0(x)}{k!}=\text{exp}(D_0(x))=L_0(x)\eeq
Note that if we are given $D(x;\hbar)$, we may always reconstruct $L_0(x)$ uniquely. On the contrary, when we are given $L(x;\hbar)$, the reconstruction of $D_0(x)$ requires to take the logarithm of the matrix $L_0$ that may be ambiguous. From equation \eqref{TRDD}, we must have $D(x;\hbar)\in \mathfrak{s}l_2(\mathbb{C})$, so that we must have $\Tr(D_0(x))=0$. Note also that when $L(x;\hbar)\in SL_2(\mathbb{C})$ and $D(x;\hbar)\in \mathfrak{s}l_2(\mathbb{C})$, we always have:
\bea \det(yI_2-D_0(x))&=&y^2-\left(\cosh^{-1}\left(\frac{\Tr L_0(x)}{2}\right)\right)^2\cr
&=&y^2-\left(\ln\left(\frac{\Tr L_0(x)}{2}+\frac{1}{2}\sqrt{(\Tr L_0(x))^2 -4}\right)\right)^2\cr
&&\eea
In particular, in the setting of \cite{Deter,BBEnew}, the characteristic polynomial of $D_0$ provides the classical spectral curve and we recover that in the $\mathbb{P}^1$ case, the classical spectral curve is given by \eqref{ClassicalCurve}. Note that the correspondence between $L_0$ and $D_0$ is not affected by the choice of $\lambda$ in the general setting of definition \ref{GeneralLSys}.

\subsection{Application to $\mathbb{P}^1$\label{AppliP11}}

General formulas \eqref{OtherEntries} and \eqref{ODEDk} simplify a lot in the case of the system \eqref{DiffSystem}. In particular since $(L_0(x))_{1,2}=-1$, it does not introduce any singularity in the induction process described in Proposition \ref{PropSing}. This is also the case for higher orders $(L_i(x))_{i\geq 1}$ that are either vanishing (for $i\geq 2$) or constant ($L_1(x)=\text{diag}(\frac{1}{2},0)$). Consequently, matrices $(D_k(x))_{k\geq 1}$ may only have singularities at points corresponding to singularities of $D_0(x)$ that may be determined using \eqref{Expo}. We obtain the following theorem: 

\begin{theorem}\label{PoleD} By induction using \eqref{OtherEntries} and \eqref{ODEDk}, we construct all matrices $\left(D_{k}(x)\right)_{k\geq 0}$ so that the $\hbar$-difference system $\delta_\hbar\Psi(x;\hbar)=L(x;\hbar)\Psi(x;\hbar)$ with $L(x;\hbar)$ given by \eqref{DiffSystem} is compatible with the differential system $\hbar \frac{d}{dx} \Psi(x;\hbar)=D(x;\hbar) \Psi(x,\hbar)$. Initialization is made by \eqref{Expo}:
\beaa D_0(x)&=&\begin{pmatrix}-\frac{x}{2\sqrt{x^2-4}}\ln\left(\frac{x+\sqrt{x^2-4}}{x-\sqrt{x^2-4}}\right)&\frac{1}{\sqrt{x^2-4}}\ln\left(\frac{x-\sqrt{x^2-4}}{x+\sqrt{x^2-4}}\right) \\\frac{1}{\sqrt{x^2-4}}\ln\left(\frac{x+\sqrt{x^2-4}}{x-\sqrt{x^2-4}}\right)& -\frac{x}{2\sqrt{x^2-4}}\ln\left(\frac{x-\sqrt{x^2-4}}{x+\sqrt{x^2-4}}\right)
\end{pmatrix}\cr
D_0(z)&\overset{\text{def}}{=}&D_0(x(z))= \begin{pmatrix}-\frac{z^2+1}{2(z^2-1)}\ln (z^2)& -\frac{z}{z^2-1}\ln(z^2)\\
\frac{z}{z^2-1}\ln(z^2)& \frac{z^2+1}{2(z^2-1)}\ln(z^2)
\end{pmatrix} \text{ with } x(z)=z+\frac{1}{z} \eeaa
In particular, we observe that $z\mapsto D_0(z)$ is regular at the branchpoints $z=\pm 1$ (or equivalently $x\mapsto D_0(x)$ is regular at $x=\pm 2$). Using Proposition \ref{PropSing}, we obtain that for all $k\geq 0$: $z\mapsto D_k(z)$ may only have singularities at $z=0$ or $z=\infty$. Note also that the characteristic polynomial of $D_0$ recovers the spectral curve \eqref{ClassicalCurve}. 
\end{theorem}

\begin{remark} The fact that the matrix $D(x;\hbar)$ is regular at the branchpoints may appear surprising since the functions $(S_k(z))_{k\geq 0}$ appearing in the expansions of the wave functions $\psi(x;\hbar)$ and $\phi(x;\hbar)$ have singularities at these points. Thus, it may appear as a surprise that the combinations appearing in $D(x;\hbar)$ are precisely those for which the singularities cancel. However, at the level of operators, this aspect appears quite natural. Indeed, at the level of operator, we formally have $\delta_\hbar= \text{exp}(\hbar \frac{d}{dx})$. Hence $L(x;\hbar)$ may be seen as the exponentiated counterpart of $D(x;\hbar)$.Equivalently, $D(x;\hbar)$ may be seen as the logarithmic version of $L(x;\hbar)$. In particular, this point of view is correct for $L_0(x)$ and $D_0(x)$. By induction (Proposition \ref{PropSing} in our case) it seems legitimate that the next orders $(D_k(x))_{k\geq 1}$ should have the same singularities as $D_0(x)$ (at least when the singularities of $\left(L_k(x)\right)_{k\geq 1}$ belong to those of $L_0(x)$). However, for a given $x\in \bar{\mathbb{C}}$, defining $\ln L_0(x)$ may not be possible. Indeed, the logarithm of a matrix is well-defined only if the matrix is invertible. Usually, in order to compute the logarithm, one diagonalizes the matrix and take the complex logarithm of the eigenvalues. Thus, two distinct problems may arise during the procedure:
\begin{itemize}\item At some values $x\in\bar{\mathbb{C}}$, some eigenvalues may coincide in such a way that the matrix cannot be diagonalized. This may only happen at the branchpoints of the classical spectral curve. However, note that if the eigenvalues are non-zero, then the logarithm remains well-defined. In fact, theorem \ref{PoleD} shows that $D_0(x)$ and more generally all $\left(D_k(x)\right)_{k\geq 0}$ are perfectly well-defined and regular at these points.
\item At some values $x\in\bar{\mathbb{C}}$, some eigenvalues may vanish, and thus the matrix is no longer invertible. In that case, the logarithm is ill-defined and we expect $D(x;\hbar)$ to be singular at those points. In our case, the eigenvalues of $L_0(x)$ are given by $\frac{1}{2}\left(-x\pm\sqrt{x^2-4}\right)$, i.e. $\frac{1}{z}$ or $z$ in the $z$-variable. In other words, $z=0$ and $z=\infty$ (i.e. $x=\infty$) are the values for which $D_0(x)$ is likely to be singular. In fact, this is precisely the result of theorem \ref{PoleD}.
\end{itemize} 
\end{remark}

\section{Constructing the matrix $M(x;\hbar)$\label{MMSection}}
\subsection{General case}

Following the works of \cite{Deter,BBEnew}, the natural next step is to define the $M(x;\hbar)$ matrix that is used in the determinantal formulas of \cite{Deter}. Thus, we introduce (we omit the index $\lambda$ but the general setting with a generic $\lambda$ applies everywhere in this section if one replace $(\Psi,L,M,D)\to(\Psi_\lambda,L_\lambda,M_\lambda,D_\lambda)$):
\beq \label{DefM}M(x;\hbar)=\Psi(x;\hbar)\begin{pmatrix} 1&0\\0&0\end{pmatrix} \Psi^{-1}(x,\hbar)\eeq
Using the two compatible systems \eqref{Compat}, we obtain that it must satisfy the equations:
\bea \label{EqM}\hbar \frac{d}{dx} M(x;\hbar)&=&\left[D(x;\hbar),M(x;\hbar)\right]\cr
\delta_\hbar M(x;\hbar)=M(x+\hbar;\hbar)&=&L(x;\hbar)M(x;\hbar)L(x;\hbar)^{-1}
\eea
The first equation is standard in the theory of differential systems while the second one is just the exponentiated form of the first one. Technically, since $L(x;\hbar)$ and $D(x;\hbar)$ can be reconstructed from one another, the former two equations are completely equivalent. In the theory of integrable systems developed in \cite{Deter,BBEnew}, one usually has access only to the first equation which is in general not sufficient to prove the topological type property. For example, in the context of Painlev\'{e} equations, an additional time-differential equation defining the Lax pair is crucial to prove the topological type property. As we will see below, in our case, the second equation involving $L(x;\hbar)$ is also of great importance since it allows to determine $M(x;\hbar)$ recursively in a very simple way. We first observe that by definition of $\Psi(x;\hbar)$ (whose determinant is set to $1$ from Proposition \ref{PropDet}), the matrix $M(x;\hbar)$ admits a formal series expansion in $\hbar$:
\beq \label{Mseries} M(x;\hbar)=\sum_{k=0}^\infty M_k(x)\hbar^k\eeq
Moreover, from its definition we have:
\beq \label{PropM} M(x;\hbar)^2=M(x;\hbar) \,\,,\,\, \Tr M(x;\hbar)=1\,\,,\,\, \det M(x;\hbar)=0\eeq
We note here some similarities with Lemma $1.2.1$ and Definition $1.2.2$ of \cite{DY2} as well as some formulas presented in \cite{NS}. Using the fact that $M(x+\hbar)=\underset{k=0}{\overset{\infty}{\sum}} \frac{1}{k!}\left(\frac{d^k}{dx^k}M(x;\hbar)\right)\hbar^k$, the second equation of \eqref{EqM} is reduced to:
\bea\label{SysM} 0&=&\left[M_0(x),L_0(x)\right] \text{ and for all } k\geq 0\,:\cr
\left[M_{k+1}(x),L_0(x)\right]&=&\sum_{m=0}^k L_{k+1-m}(x)M_m(x)-\sum_{\substack{i+j+m=k+1\\ m\leq k}}\frac{1}{j!}\left(\frac{d^j}{dx^j}M_m(x)\right)L_i(x)\cr
&&
\eea  
In particular, if $L(x;\hbar)=L_0(x)+L_1(x)\hbar$, the previous equations simplify into:
\bea \label{SysMSimplified} 0&=&\left[M_0(x),L_0(x)\right] \cr
\left[M_{k+1}(x),L_0(x)\right]&=&[L_1(x),M_k(x)]-\sum_{j=1}^{k+1}\frac{1}{j!}\left(\frac{d^j}{dx^j}M_{k+1-j}(x)\right)L_0(x)\cr
&&-\sum_{j=1}^{k}\frac{1}{j!}\left(\frac{d^j}{dx^j}M_{k-j}(x)\right)L_1(x) \,\,,\forall\, k\geq 0
\eea

Similarly to the differential case, expressing the commutator $[M_{k+1}(x),L_0(x)]$ in terms of lower orders from \eqref{SysM} only provides two linearly independent equations on the entries of $M_{k+1}(x)$. Nevertheless, conditions $\Tr M(x;\hbar)=1$ and $\det M(x;\hbar)=0$ (or in higher dimensions, the fact that $M(x;\hbar)^2=M(x;\hbar)$) are sufficient to provide enough linearly independent equations on the entries. We find:

\begin{theorem}\label{GeneralRecM} The first order $M_0(x)$ is given by:
\beqq M_0(x)=\begin{pmatrix}\frac{1}{2}+\frac{(L_0(x))_{1,1}-(L_0(x))_{2,2}}{\sqrt{(\Tr L_0(x))^2-4\det L_0(x)}} &\frac{(L_0(x))_{1,2}}{\sqrt{(\Tr L_0(x))^2-4\det L_0(x)}}\\ \frac{(L_0(x))_{2,1}}{\sqrt{(\Tr L_0(x))^2-4\det L_0(x)}}& \frac{1}{2}-\frac{(L_0(x))_{1,1}-(L_0(x))_{2,2}}{\sqrt{(\Tr L_0(x))^2-4\det L_0(x)}}
 \end{pmatrix}\eeqq
Higher orders may be obtained by recursion from the resolution of the linear system (along with $(M_{k+1}(x))_{2,2}=-(M_{k+1}(x))_{1,1}$) for $k\geq 0$:
\small{\beaa&&\begin{pmatrix} 0&(L_0(x))_{2,1}& -(L_0(x))_{1,2}\\
2(L_0(x))_{1,2}& (L_0(x))_{2,2}-(L_0(x))_{1,1}&0\\
(L_0(x))_{2,2}-(L_0(x))_{1,1}&-(L_0(x))_{2,1} &-(L_0(x))_{1,2}
\end{pmatrix}
\begin{pmatrix}(M_{k+1}(x))_{1,1}\\(M_{k+1}(x))_{1,2}\\(M_{k+1}(x))_{2,1}
\end{pmatrix}=\cr
&&\begin{pmatrix}\left(\underset{m=0}{\overset{k}{\sum}} L_{k+1-m}(x)M_m(x)-\underset{{\substack{i+j+m=k+1\\ m\leq k}}}{\sum}\frac{1}{j!}\left(\frac{d^j}{dx^j}M_m(x)\right)L_i(x)\right)_{1,1}\\
 \left(\underset{m=0}{\overset{k}{\sum}} L_{k+1-m}(x)M_m(x)-\underset{{\substack{i+j+m=k+1\\ m\leq k}}}{\sum}\frac{1}{j!}\left(\frac{d^j}{dx^j}M_m(x)\right)L_i(x)\right)_{1,2}\\
 \sqrt{(\Tr L_0(x))^2-4\det L_0(x)}\underset{j=1}{\overset{k}{\sum}}\left((M_j(x))_{1,1}(M_{k+1-j}(x))_{1,1}+(M_j(x))_{1,2}(M_{k+1-j}(x))_{2,1}\right)\end{pmatrix}\cr
\eeaa}\normalsize{}
Note that the determinant of the recursion matrix is given by \beqq(L_0(x))_{1,2}(x)\left((\Tr L_0(x))^2-4\det L_0(x)\right)\eeqq
Therefore, we can invert the matrix and obtain:
\footnotesize{\beaa&&\begin{pmatrix}(M_{k+1})_{1,1}\\(M_{k+1})_{1,2}\\(M_{k+1})_{1,2}\end{pmatrix}=\frac{1}{(\Tr L_0)^2-4\det L_0}\cr
&&\begin{pmatrix} (L_0)_{1,1}-(L_0)_{2,2}& 2(L_0)_{2,1}& (L_0)_{2,2}-(L_0)_{1,1}\\
2(L_0)_{1,2}&(L_0)_{2,2}-(L_0)_{1,1}&-2(L_0)_{1,2}\\
2(L_0(x))_{2,1}-\frac{(\Tr L_0)^2-4\det L_0}{(L_0)_{1,2}}&\frac{((L_0)_{2,2}-(L_0)_{1,1})(L_0)_{2,1}}{(L_0)_{1,2}}&-2(L_0)_{2,1}
\end{pmatrix}\cr
&&\begin{pmatrix}\left(\underset{m=0}{\overset{k}{\sum}} L_{k+1-m}(x)M_m(x)-\underset{{\substack{i+j+m=k+1\\ m\leq k}}}{\sum}\frac{1}{j!}\left(\frac{d^j}{dx^j}M_m(x)\right)L_i(x)\right)_{1,1}\\
 \left(\underset{m=0}{\overset{k}{\sum}} L_{k+1-m}(x)M_m(x)-\underset{{\substack{i+j+m=k+1\\ m\leq k}}}{\sum}\frac{1}{j!}\left(\frac{d^j}{dx^j}M_m(x)\right)L_i(x)\right)_{1,2}\\
 \sqrt{(\Tr L_0(x))^2-4\det L_0(x)}\underset{j=1}{\overset{k}{\sum}}\left((M_j(x))_{1,1}(M_{k+1-j}(x))_{1,1}+(M_j(x))_{1,2}(M_{k+1-j}(x))_{2,1}\right)\end{pmatrix}\cr
\eeaa}\normalsize{}
\end{theorem}

\begin{remark} We observe that the recursive construction of the matrix $M(x;\hbar)$ may only be performed using the matrix $L(x;\hbar)$ but not directly the matrix $D(x;\hbar)$. Moreover, we note that the reconstruction formulas given in theorem \ref{GeneralRecM} appear simpler than the reconstruction formulas obtained from the differential system $\hbar \frac{d}{dx} M(x;\hbar)=D(x;\hbar)M(x;\hbar)$ (see \cite{Deter,BBEnew,P2,Painleve} for the explicit formulas). As we have seen above, both systems are equivalent but, depending on the problem, one of the formulations may be easier to handle than the other one (in particular if one the matrices $D(x;\hbar)$ or $L(x;\hbar)$ has a simple dependence in $x$). 
\end{remark}

\subsection{Application to $\mathbb{P}^1$ \label{SectionMExplicit}}

Applying the general formulas of theorem \ref{GeneralRecM} in the $\mathbb{P}^1$ case gives:
\begin{corollary}\label{ExpressionM}
For the $\hbar$-difference system given by \eqref{DiffSystem}, we have:
\beq \label{M0} M_0(x)=\begin{pmatrix}\frac{1}{2}+\frac{x}{2\sqrt{x^2-4}}& -\frac{1}{\sqrt{x^2-4}}\\ \frac{1}{\sqrt{x^2-4}}& \frac{1}{2}-\frac{x}{2\sqrt{x^2-4}}
\end{pmatrix}\eeq
and the recursion $\forall \, k\geq 0$:
\footnotesize{\bea\label{Mrec}
\begin{pmatrix} (M_{k+1}(x))_{1,1}\\ (M_{k+1}(x))_{1,2}\\ (M_{k+1}(x))_{2,1}\end{pmatrix}&=&\frac{1}{x^2-4}\begin{pmatrix}x&2&-x\\
-2&-x&2\\
x^2-2&x&-2\end{pmatrix}\cr
&&
\begin{pmatrix}-\underset{j=1}{\overset{k+1}{\sum}}\frac{1}{j!}\left(x\frac{d^j}{dx^j}(M_{k+1-j}(x))_{1,1}+\frac{d^j}{dx^j}(M_{k+1-j}(x))_{1,2}\right)-\frac{1}{2}\underset{j=1}{\overset{k}{\sum}}\frac{1}{j!}\frac{d^j}{dx^j}(M_{k-j}(x))_{1,1}\\
 \frac{1}{2}(M_k)_{1,2}+\underset{j=1}{\overset{k+1}{\sum}}\frac{1}{j!}\left(\frac{d^j}{dx^j}(M_{k+1-j}(x))_{1,1}\right)\\
 \sqrt{x^2-4}\underset{j=1}{\overset{k}{\sum}}\left((M_j(x))_{1,1}(M_{k+1-j}(x))_{1,1}+(M_j(x))_{1,2}(M_{k+1-j}(x))_{2,1}\right)\end{pmatrix}\cr
&&
\eea}\normalsize{}
\end{corollary}

Hence, we obtain the following proposition regarding the singularities structure of the matrices $\left(M_k(x)\right)_{k\geq 0}$ in the $\mathbb{P}^1$ case:

\begin{proposition}\label{PoleStructure}For all $k\geq 0$, the functions $z\mapsto M_k(x(z))$ are rational functions that may only have poles at the branchpoints $z=\pm 1$ or at $z=\infty$. Moreover, at $x\to \infty$ we have:
\beqq M_0(x)=\begin{pmatrix}1&0\\0&0\end{pmatrix}+O\left(\frac{1}{x^2}\right) \text{ and } M_k(x)=O\left(\frac{1}{x^2}\right) \text{ for } k\geq 1\eeqq
\end{proposition}

\proof{The proof is straightforward from \eqref{M0} and \eqref{Mrec}. We first observe that $M_0(x)$ has indeed the correct behavior and that the claim is also correct for $M_1(x)$ that is given by:
\beq M_1(x)=\frac{1}{(x^2-4)^{\frac{3}{2}}}\begin{pmatrix}1&0\\x&-1
\end{pmatrix}
\eeq
Then by induction, if we assume the proposition to be correct for all $j\leq k$, we may use \eqref{Mrec} to prove the right behavior of $M_{k+1}(x)$. Indeed, we first observe that $M_0$ is only involved through its derivatives, hence the factor $\text{diag}(1,0)$ will not play any role. Then the result is obvious from a direct analysis of the various quantities at infinity.
}

\subsubsection{Computations of the first orders of $M(x;\hbar)$\label{SecSec}}
Implementing the recursion to determine $(M_k(x))_{k\geq 0}$ is straightforward. In order to compare the formal expansion at $x\to \infty$ with those proposed in conjecture \eqref{ConjectureM}, we provide below the first orders of the matrix $M(x;\hbar)$:
\beaa M_0(x)&=&\left( \begin {array}{cc} \frac{1}{2}+{\frac {x}{2\sqrt {{x}^{2}-4}}}&-{\frac {1}{\sqrt {{x}^{2}-4}}}\\\noalign{\medskip}{\frac {1}{\sqrt {{x}
^{2}-4}}}&\frac{1}{2}-\,{\frac {x}{2\sqrt {{x}^{2}-4}}}\end {array} \right)\cr
M_1(x)&=&\left( \begin {array}{cc} 0&\,{\frac {x}{ 2\left( {x}^{2}-4 \right) ^{3/2}}}\\\noalign{\medskip}\,{\frac {x}{ 2\left( {x}^{2}-4
 \right) ^{3/2}}}&0\end {array} \right)\cr
M_2(x)&=&\left( \begin {array}{cc} \,{\frac {x \left( {x}^{2}+16 \right) }{ 4\left( {x}^{2}-4 \right) ^{7/2}}}&-\,{\frac {{x}^{2} \left( {x}^{2
}+6 \right) }{ 4\left( {x}^{2}-4 \right) ^{7/2}}}\\\noalign{\medskip}{\frac {{x}^{2} \left( {x}^{2}+6 \right) }{ 4\left( {x}^{2}-4
 \right) ^{7/2}}}&-\,{\frac {x \left( {x}^{2}+16 \right) }{4 \left( 
{x}^{2}-4 \right) ^{7/2}}}\end {array} \right)\cr
M_3(x)&=&\left( \begin {array}{cc} 0&\,{\frac {x \left( {x}^{4}+42\,{x}^{2}+96 \right) }{8 \left( {x}^{2}-4 \right) ^{9/2}}}\\\noalign{\medskip}\,{\frac {x \left( {x}^{4}+42\,{x}^{2}+96 \right) }{ 8\left( {x}^{2}-4
 \right) ^{9/2}}}&0\end {array} \right)\cr 
M_4(x)&=&\left( \begin {array}{cc} \,{\frac {x \left( {x}^{6}+247\,{x}^{4}+2848\,{x}^{2}+3072 \right) }{16 \left( {x}^{2}-4 \right) ^{13/2}}}&-\,{\frac {{x}^{2} \left( {x}^{6}+156\,{x}^{4}+1350\,{x}^{2}+1280
 \right) }{16 \left( {x}^{2}-4 \right) ^{13/2}}}\\\noalign{\medskip}
\,{\frac {{x}^{2} \left( {x}^{6}+156\,{x}^{4}+1350\,{x}^{2}+1280
 \right) }{16 \left( {x}^{2}-4 \right) ^{13/2}}}&-\,{\frac {x
 \left( {x}^{6}+247\,{x}^{4}+2848\,{x}^{2}+3072 \right) }{16 \left( {x}^
{2}-4 \right) ^{13/2}}}\end {array} \right)\cr
M_5(x)&=& \left( \begin {array}{cc} 0&\,{\frac {x \left( 30720+52160\,{x}^{
2}+12990\,{x}^{4}+{x}^{8}+516\,{x}^{6} \right) }{32 \left( {x}^{2}-4
 \right) ^{15/2}}}\\\noalign{\medskip}\,{\frac {x \left( 30720+
52160\,{x}^{2}+12990\,{x}^{4}+{x}^{8}+516\,{x}^{6} \right) }{32 \left( {
x}^{2}-4 \right) ^{15/2}}}&0\end {array} \right)\cr
\eeaa 

\subsubsection{Some symmetries for the entries of $M(x;\hbar)$\label{Symme}}
A direct computation from the definition shows that:
\beq\label{MM} M(x;\hbar)=\begin{pmatrix}\psi(x+\frac{\hbar}{2};\hbar)\phi(x-\frac{\hbar}{2};\hbar)& -\psi(x+\frac{\hbar}{2};\hbar)\phi(x+\frac{\hbar}{2};\hbar)\\ \psi(x-\frac{\hbar}{2};\hbar)\phi(x-\frac{\hbar}{2};\hbar)& -\phi(x+\frac{\hbar}{2};\hbar)\psi(x-\frac{\hbar}{2};\hbar)
\end{pmatrix}\eeq
Thus, since $\psi(x;\hbar)\phi(x;\hbar)$ admits a WKB expansion with only even powers of $\hbar$, we get that:
\beq (M_k(x;\hbar))_{2,1}=(-1)^{k+1}(M_k(x))_{1,2}\eeq
A similar observation for $\psi(x+\frac{1}{2}\hbar;\hbar)\phi(x-\frac{1}{2}\hbar;\hbar)$ shows that it WKB expansion may only involve even powers of $\hbar$ so that: $(M_{2k+1})_{1,1}=0$ for all $k\geq 0$. Moreover, from $\Tr M_k(x)=0$ for $k\geq 1$, we conclude that the only non-trivial entries of $\left(M_k(x)\right)_{k\geq 1}$ are:
\beqq (M_{2k})_{1,1}, (M_{2k})_{1,2} \text{ and } (M_{2k+1})_{1,2}\eeqq
Eventually, we note that these symmetries correspond exactly to those proposed in conjecture \eqref{ConjectureM}.

\subsubsection{Reduced system for $(M_{2k})_{1,1}$, $(M_{2k})_{1,2}$ and $(M_{2k+1})_{1,2}$}
Using the symmetries presented in section \ref{Symme}, we may rewrite a reduced recursive system for the entries $(M_{2k})_{1,1}$, $(M_{2k})_{1,2}$ and $(M_{2k+1})_{1,2}$ from corollary \ref{ExpressionM}. We find:
\small{\bea\label{ReducedSystem} (M_{2k+1})_{1,1}&=&\frac{1}{x^2-4}\Big[ (2-x)^2\sum_{l=0}^{k} \frac{1}{(2l+1)!}\frac{d^{2l+1}}{dx^{2l+1}}(M_{2k-2l})_{1,1}-\frac{x}{2}\sum_{l=1}^k\frac{1}{(2l)!}\frac{d^{2l}}{dx^{2l}}(M_{2k-2l})_{1,1}\cr
&&-x\sum_{l=1}^k \frac{1}{(2l)!}\frac{d^{2l}}{dx^{2l}}(M_{2k-2l+1})_{1,2} -x\sum_{l=0}^k \frac{1}{(2l+1)!}\frac{d^{2l+1}}{dx^{2l+1}}(M_{2k-2l})_{1,2}\Big]\cr
(M_{2k})_{1,1}&=&\frac{1}{x^2-4}\Big[(2-x^2)\sum_{l=1}^k \frac{1}{(2l)!}\frac{d^{2l}}{dx^{2l}}(M_{2k-2l})_{1,1}-\frac{x}{2}\sum_{l=1}^k\frac{1}{(2l-1)!}\frac{d^{2l-1}}{dx^{2l-1}}(M_{2k-2l})_{1,1}\cr
&&-x\sum_{l=1}^k\frac{1}{(2l)!}\frac{d^{2l}}{dx^{2l}}(M_{2k-2l})_{1,2}-x\sum_{l=1}^k\frac{1}{(2l-1)!}\frac{d^{2l-1}}{dx^{2l-1}}(M_{2k-2l+1})_{1,2}\cr
&&+(M_{2k-1})_{1,2}\Big]\cr
&&-x\sqrt{x^2-4}\Big[\sum_{l=1}^{k-1}\left((M_{2l})_{1,1}(M_{2k-2l})_{1,1}-(M_{2l})_{1,2}(M_{2k-2l})_{1,2}\right)\cr
&&+\sum_{l=1}^k(M_{2l-1})_{1,2}(M_{2k-2l+1})_{1,2}\Big]\cr
(M_{2k})_{1,2}&=&\frac{1}{x^2-4}\Big[x\sum_{l=1}^k\frac{1}{(2l)!}\frac{d^{2l}}{dx^{2l}}(M_{2k-2l})_{1,1}+\sum_{l=1}^k \frac{1}{(2l-1)!}\frac{d^{2l-1}}{dx^{2l-1}}(M_{2k-2l})_{1,1}\cr
&&+2\sum_{l=1}^k\frac{1}{(2l)!}\frac{d^{2l}}{dx^{2l}}(M_{2k-2l})_{1,2}+2\sum_{l=1}^k\frac{1}{(2l-1)!}\frac{d^{2l-1}}{dx^{2l-1}}(M_{2k-2l+1})_{1,2}\cr
&&-\frac{x}{2}(M_{2k-1})_{1,2}\Big]\cr
&&+\frac{2}{\sqrt{x^2-4}}\Big[\sum_{l=1}^{k-1}\left((M_{2l})_{1,1}(M_{2k-2l})_{1,1}-(M_{2l})_{1,2}(M_{2k-2l})_{1,2}\right)\cr
&&+\sum_{l=1}^k(M_{2l-1})_{1,2}(M_{2k-2l+1})_{1,2}\Big]\cr
&&
\eea}\normalsize{} 
 
By induction, it is then straightforward to prove that:
\bea (M_{2k})_{1,1}&=&\frac{xQ_{4k-2}(x)}{(x^2-4)^{\frac{6k+1}{2}}} \,\,,\,\, (M_{2k})_{1,2}=\frac{x^2R_{4k-2}(x)}{(x^-4)^{\frac{6k+1}{2}}}\cr
(M_{2k-1})_{1,2}&=&\frac{xT_{4k-4}(x)}{(x^2-4)^{\frac{6k-3}{2}}} \,\,,\,\, \forall\, k\geq 1
\eea
 where $Q_{2k-2}$, $R_{2k-2}$ and $T_{4k}$ are polynomials of degree $2k-2$, $2k-2$ and $4k$ respectively.

\section{Determinantal formulas\label{DeterSection}}

\subsection{Definition of the correlation functions}

In \cite{Deter}, the authors proposed to associate to any $2\times 2$ differential system of the form $\hbar \frac{d}{dx} \Psi(x;\hbar)=D(x;\hbar)\Psi(x;\hbar)$ some correlation functions $W_n(x_1,\dots,x_n;\hbar)$ (with $n\geq 1$) through determinantal formulas. This construction was later extended to arbitrary dimension in \cite{BBEnew} but since we will only need the case $d=2$ in this article, we will only review the simpler presentation of \cite{Deter}.
As explained in \cite{Deter}, there exist two equivalent ways to define the correlation functions. The first definition is obtained from the Christoffel-Darboux kernel which is defined from the entries of the matrix $\Psi(x;\hbar)$ (given in \eqref{Psi}). In our case, it is given by:
\beq \label{KK} 
K(x_1,x_2;\hbar)=\frac{\psi(x_1+\frac{\hbar}{2};\hbar)\phi(x_2-\frac{\hbar}{2};\hbar)-
\psi(x_1-\frac{\hbar}{2};\hbar)\phi(x_2++\frac{\hbar}{2};\hbar)}{x_1-x_2}
\eeq
Then the correlation functions are defined by:
\begin{definition}[Definition 2.3 of {\cite{Deter}}]\label{DeterForm} The (connected) correlation functions are defined by:
\beaa \label{DefCorrDarboux}
W_1(x;\hbar)&=&\left(\frac{d}{dx}\psi(x+\frac{\hbar}{2};\hbar)\right)\phi(x-\frac{\hbar}{2};\hbar)-\left(\frac{d}{dx}\psi(x-\frac{\hbar}{2};\hbar)\right)\phi(x+\frac{\hbar}{2};\hbar),\\
W_n(x_1,\dots,x_n;\hbar)&=&-\frac{\delta_{n,2}}{(x_1-x_2)^2}+(-1)^{n+1}\sum_{\sigma: \text{$n$-cycles}}\prod_{i=1}^n K(x_i,x_{\sigma(i)};\hbar) ~\quad\text{for $n \ge 2$} \eeaa where $\sigma$ is a $n$-cycle permutation.
\end{definition}
For our purpose, it is important to mention that there exists an alternative definition (the equivalence of the two definitions can be found in \cite{Deter}) in terms of the matrix $M(x;\hbar)$.
\begin{definition} The correlation functions may also be defined with the following formulas:
\beaa \label{alternative} 
W_1(x;\hbar)&=&-\frac{1}{\hbar} \Tr (D(x;\hbar)M(x;\hbar)),\\
W_2(x_1,x_2;\hbar)&=&\frac{\Tr (M(x_1;\hbar)M(x_2;\hbar)) -1}{(x_1-x_2)^2},\\
W_n(x_1,\dots,x_n;\hbar)&=& (-1)^{n+1}\Tr \sum_{\sigma: \text{$n$-cycles}} \prod_{i=1}^n \frac{M(x_{\sigma^i(1)};\hbar)}{x_{\sigma^i(1)}-x_{\sigma^{i+1}(1)}}\\
&=&\frac{(-1)^{n+1}}{n}\sum_{\sigma \in S_n} \frac{ \Tr M(x_{\sigma(1)};\hbar)\dots M(x_{\sigma(n)};\hbar)}{(x_{\sigma(1)}-x_{\sigma(2)})\dots (x_{\sigma(n-1)}-x_{\sigma(n)})(x_{\sigma(n)}-x_{\sigma(1)})} \\
&&  \text{for } n \geq 3. 
\eeaa
\end{definition}

The second definition makes the connection with the conjecture of \cite{Dubrovin} obvious. Therefore we need to prove that the correlation functions defined from the determinantal formulas identify with the Eynard-Orantin differentials computed from the topological recursion. This can be done using a sufficient set of conditions known as the topological type property \cite{BBEnew}.

\subsection{The topological type property\label{TTSection}}

In this section, we review the topological type property. We restrict ourselves to the case of genus $0$ classical spectral curve $\det(yI_d-D_0(x))=0$ for which the topological type property is simpler but we mention that a more general version can be found in \cite{BBEnew} when the genus is not vanishing.

\begin{definition}\label{TTdef}[Topological type property for genus $0$ curves] A differential system $\hbar \frac{d}{dx} \Psi(x;\hbar)=D(x;\hbar)\Psi(x;\hbar)$ where $D(x;\hbar)$ admits a formal series expansion in $\hbar$ of the form
\beqq D(x;\hbar)=\sum_{k=0}^\infty D_k(x)\hbar^k\in\mathcal{M}_d(\mathbb{C}(x)) \text{ with } \det(yI_d -D_0(x))=0 \text{ a genus 0 curve}\eeqq
is said to be of topological type if the following conditions are met:
\begin{enumerate} 
\item[(1)] \underline{Existence of a series expansion in $\hbar$}: The correlation functions admit a formal series expansion in $\hbar$ of the form:
\beaa W_1(x;\hbar)&=&\sum_{k=-1}^\infty W_1^{(k)}(x)\hbar^k\cr
W_n(x_1,\dots,x_n;\hbar)&=&\sum_{k=0}^\infty W_n^{(k)}(x_1,\dots,x_n)\hbar^k \,\,,\,\, \forall \, n\geq 2
\eeaa
\item[(2)] \underline{Parity property}: $W_{n}|_{\hbar \hspace{+.1em} \mapsto - \hbar} = (-1)^{n} W_{n}$ holds for $n \geq 1$. This is equivalent to say that the previous series expansion is even (resp. odd) when $n$ is even (resp. odd).
\item[(3)] \underline{Pole structure}: The functions $(z_1,\dots,z_n)\mapsto W_n^{(k)}(x(z_1),\dots,x(z_n);\hbar)$ are rational functions that may only have poles at the branchpoints when $(g,n) \neq (0,1), (0,2)$. Moreover:
\beqq \left(W_2^{(0)}(x_1,x_2)+\frac{1}{(x_1-x_2)^2}\right)dx_1dx_2=\frac{\Tr (M_0(x_1)M_0(x_2)) -1}{(x_1-x_2)^2}dx_1dx_2\eeqq
should identify with the normalized bi-differential $\omega_2^{(0)}$ associated to the classical spectral curve.  
\item[(4)] \underline{Leading order}: The leading order of the series expansion of the correlation function $W_n$ is at least of order $\hbar^{n-2}$.
\end{enumerate}
In fact, combining conditions $1$, $2$ and $3$ is equivalent to require that:
\beqq W_n(x_1,\dots,x_n;\hbar)=\sum_{g=0}^\infty W_n^{(n-2+2g)}(x_1,\dots,x_n)\hbar^{n-2+2g}\eeqq
\end{definition}

The topological type property is particularly useful since we have the following theorem:

\begin{theorem}[Theorem 3.1 and Corollary $4.2$ of \cite{BBEnew}]\label{thm:TT} 
If the differential system $\hbar\frac{d}{dx} \Psi(x;\hbar)=D(x;\hbar)\Psi(x;\hbar)$ has a spectral curve of genus $0$ and satisfies the topological type property \ref{TTdef}, then the differentials $W_n^{(n-2+2g)}(x_{1},\dots,x_{n})dx_1\dots dx_n$ appearing in the formal expansion of the correlation functions $W_{n}(x_{1},\dots,x_{n};\hbar)dx_1\dots dx_n$ are identical to the Eynard-Orantin differentials $\omega^{(g)}_{n}$ obtained from the application of the topological recursion to the spectral curve $\det(yI_d-D_0(x))=0$. In other words:
\beqq \label{TheoWN}
W^{(n-2+2g)}_{n}(x(z_{1}),\dots,x(z_{n}))dx(z_{1})\dots dx(z_{n}) = \omega^{(g)}_{n}(z_{1},\dots,z_{n}) \quad \text{for $g \geq 0$, $n \geq 1$}
\eeqq
where $x(z)$ is a parametrization of the spectral curve. This is also equivalent to say that:
\beqq \forall \,n\geq 1\,:\, W_n(x(z_1),\dots,x(z_n))dx(z_1)\dots dx(z_n)=\sum_{g=0}^{\infty} \omega^{(g)}_{n}(z_{1},\dots,z_{n})\hbar^{n-2+2g}\eeqq  
\end{theorem}

Therefore, a natural strategy is to prove the topological type property to obtain the identification of the correlation functions defined from the determinantal formulas with the Eynard-Orantin differentials arising from the topological recursion.

\begin{remark}A complete proof of Theorem $3.1$ of \cite{BBEnew} may be found in Appendix $B$ of the same paper. It uses ingredients from \cite{AbstractLoop,EO} but is self-contained.
\end{remark}

\subsection{Proving the topological type property for $\mathbb{P}^1$\label{ProofTT}}

In this section we prove that the differential system \eqref{Compat} arising in the Gromov-Witten invariants of $\mathbb{P}^1$ satisfies the topological type property. 

\subsubsection{Existence of a formal series in $\hbar$}
In the $\mathbb{P}^1$ case, the first point of definition \ref{TTdef} is obvious. Indeed, as we have seen above, the matrix $M(x;\hbar)$ admits a formal series expansion in $\hbar$ given by \eqref{Mseries} (which is a consequence of the fact that we have taken formal WKB solutions in $\Psi(x;\hbar)$). Consequently, from the alternative definition of the correlation functions \eqref{alternative}, we immediately get that all correlation functions $W_n(x_1,\dots,x_n)$ admit a formal series expansion in $\hbar$. Note that this also holds for $W_1(x)$ from either \eqref{DefCorrDarboux} (where one can see that the $\text{exp}(\pm\frac{1}{\hbar}S_0(x))$ terms cancel) or from \eqref{alternative} (and the fact that $D(x;\hbar)$ also admit a formal series expansion in $\hbar$).

\subsubsection{Pole structure}
In the $\mathbb{P}^1$ case, proving the third point of definition \ref{TTdef} is also straightforward. Indeed, from proposition \ref{PoleStructure} we know that for all $k\geq 0$, $z\mapsto M_k(x(z))$ is a rational function that may only have singularities at $z\in\{-1,1,\infty\}$. Hence by the alternative definition \eqref{alternative}, this property also holds for the correlation functions $W_n(x(z_1),\dots,x(z_n))$ with $n\geq 2$. Moreover, for $n\geq 2$, the behavior at $x\to \infty$ of the matrices $\left(M_k(x)\right)_{k\geq 0}$ given by proposition \ref{PoleStructure}, combined with the denominators (where each $x_i$'s appears twice) proves that $W_n(x(z_1),\dots,x(z_n))$ is $O\left(\frac{1}{z_i^2}\right)$ for all $1\leq i\leq n$. In particular, it has no pole at $z_i\to\infty$ and thus may only have poles at $z_i=\pm 1$ as requested. Proving the behavior of $W_1(x)$ is more complicated. Indeed, its definition involves the matrix $D(x;\hbar)$ whose construction by recursion is rather complicated. We prefer instead the use the following observation: from its definition \eqref{DefCorrDarboux}, the application of the operator $\delta_\hbar$ on $W_1(x;\hbar)$ gives:
\beq \label{EqW1}\delta_\hbar W_1(x;\hbar)=W_1(x;\hbar)-M_{1,1}(x;\hbar)\eeq
 
Reminding that $W_1(x)=\underset{k=-1}{\overset{\infty}{\sum}} W_{1}^{(k)}(x)\hbar^k$ (i.e. the series expansion starts at $O(\hbar^{-1})$), equation \eqref{EqW1} is equivalent to:
\beq \sum_{k=-1}^\infty\sum_{j=0}^\infty \frac{1}{j!}\left(\frac{d^j}{dx^j}W_1^{(k)}(x)\right)\hbar^{k+j}=\sum_{k=-1}^\infty W_1^{(k)}(x)\hbar^{k}-\sum_{k=0}^\infty(M_k(x))_{1,1} \hbar^k\eeq
This system can be solved order by order in $\hbar$:
\bea \label{RecursiveSystem}\frac{d}{dx}W_1^{(-1)}(x)&=&-(M_0(x))_{1,1}\cr
\frac{d}{dx}W_1^{(0)}(x)&=&0\cr
\frac{d}{dx}W_1^{(k)}(x)&=&-\sum_{j=2}^{k+1} \frac{1}{j!}\frac{d^j}{dx^j}W_1^{(k+1-j)}(x) -(M_{k+1}(x))_{1,1} \,,\,\forall k\geq 1\cr
\eea
Using the properties on the matrices $(M_{k}(x))_{k\geq 0}$ of section \ref{SectionMExplicit}, we find by taking $W_1^{(k)}(x)=\int_\infty^x \frac{d}{dx'}W_1^{(k)}(x')dx'$ with a trivial recursion that:
\beq \forall \, k\geq 0: W_1^{(k)}(x)\underset{x\to\infty}{=}O\left(\frac{1}{x^2}\right)\eeq 
Therefore, from proposition \ref{PoleStructure}, we obtain that for all $k\geq 1$, the one-form $z\mapsto W_1^{(k)}(x(z))dz$ may only have singularities at the branchpoints and that these singularities may only be poles. Hence we finally have the following theorem:  

\begin{theorem}\label{PoleStructureDiffSystem} The correlation functions $(z_1,\dots,z_n)\mapsto (W_n^{(k)}(x(z_1),\dots, x(z_n)))_{n\geq 1, k\geq 0}$ with $(n,k)\neq (0,1)$ defined from the $\hbar$-difference system \eqref{DiffSystem} are rational functions that may only have poles at the branchpoints $z_i=\pm 1$. In particular we have for all $(n,k)\neq(1,0)$:
\beqq  \forall\, 1\leq i\leq n\,:\, W_n^{(k)}(x_1,\dots, x_n)\overset{x_i\to \infty}{=}O\left(\frac{1}{x_i^2}\right)\eeqq
Moreover, from an explicit computation we have using \eqref{M0}:
\beaa &&\left(W_2^{(0)}(x(z_1),x(z_2))+\frac{1}{(x(z_1)-x(z_2))^2}\right)dx(z_1)dx(z_2)\cr
&=&\frac{\Tr(M_0(x(z_1))M_0(x(z_2)))}{(x(z_1)-x(z_1))^2}dx(z_1)dx(z_2)\cr
&=&\frac{dz_1dz_2}{(z_1-z_2)^2}\eeaa
which is the standard normalized bi-differential $\omega_2^{(0)}$ associated to a genus $0$ spectral curve.
\end{theorem}

\subsubsection{Parity property \label{SectionParity}}
In order to prove the second property of the topological type property, we use a sufficient condition proposed in \cite{BBEnew}:

\begin{proposition}[Proposition $3.3$ of \cite{BBEnew}]\label{PropSuf} Let us denote by $\dagger$ the operator switching $\hbar$ into $-\hbar$. If there exists an invertible matrix $\Gamma$ independent of $x$ such that
\beqq\label{Gamma} 
\Gamma^{-1} D^t(x;\hbar)\Gamma=D^\dagger(x;\hbar),\eeqq
then the correlation functions $W_{n}(x_1,\dots,x_n;\hbar)$ satisfy
\beqq \forall \, n\geq 1\,:\, W_{n}^{\dagger} = (-1)^{n} W_{n}.\eeqq 
\end{proposition}

This sufficient condition was used successfully in \cite{P2,Painleve}. To our knowledge, there is no known case that satisfy the parity property but that does not satisfy proposition \ref{PropSuf}. As one can see, the main difficulty compared to \cite{P2,Painleve} is that the previous condition involves the matrix $D(x;\hbar)$ rather than the matrix $L(x;\hbar)$ that would have been more natural in our context. The key element in the last proposition is to correctly understand the action of the operator $\,^\dagger$ changing $\hbar\to -\hbar$. Since our quantum curve \eqref{QuantumCurve} is our starting point and is invariant under this transformation, the action of the operator $\,^\dagger$ on the matrix $\Psi(x;\hbar)$ is simple: it roughly exchanges the role of $\psi(x;\hbar)$ and $\phi(x;\hbar)$. More precisely, if we take into account that we have imposed $\det\Psi(x;\hbar)=1$ (proposition \ref{PropDet}), we have (we omit to write $;\hbar$ in all quantities for clarity):
\beq \Psi(x)=\begin{pmatrix}\psi(x+\frac{\hbar}{2})&\phi(x+\frac{\hbar}{2})\\ \psi(x-\frac{\hbar}{2})&\phi(x-\frac{\hbar}{2})\end{pmatrix} \,\Rightarrow\, \Psi^\dagger(x)=\begin{pmatrix} \phi(x+\frac{\hbar}{2})&-\psi(x+\frac{\hbar}{2})\\ \phi(x-\frac{\hbar}{2})&-\psi(x-\frac{\hbar}{2})\end{pmatrix} \eeq
Observe in particular that the last identity is equivalent to say that:
\beq \label{PsiDagger} \Psi^\dagger(x)=\begin{pmatrix} 0&-1\\1&0\end{pmatrix} \left(\Psi(x)^{-1}\right)^t\overset{\text{def}}{=}A \left(\Psi(x)^{-1}\right)^t\eeq
It is then a simple computation to prove that the last identity implies:
\beq D^\dagger(x)=-\hbar\left(\frac{d}{dx}\Psi^\dagger(x)\right)\left(\Psi^\dagger(x)\right)^{-1}=A D(x)^t A^{-1}\eeq
Thus, we take:
\beq\Gamma=A^{-1}=\begin{pmatrix} 0&1\\-1&0\end{pmatrix}\eeq
to satisfy proposition \ref{PropSuf}. For completeness, we now compute $L^\dagger(x)$. Acting on $\Psi^\dagger$ with the operator $\delta_{-\hbar}$ gives:
\beq L^\dagger(x)=A L^t(x-\hbar) A^{-1}=\Gamma^{-1} \eeq
Hence we have:
\beq L^\dagger(x)=\Gamma^{-1} \begin{pmatrix} x-\frac{\hbar}{2} & 1\\ -1&0 \end{pmatrix} \Gamma\eeq

\begin{theorem}\label{ParityDiffSystem} The correlation functions $(z_1,\dots,z_n)\mapsto (W_n(x(z_1),\dots, x(z_n)))_{n\geq 1}$ defined from the system \eqref{DiffSystem} satisfy the parity property. 
\end{theorem}

\begin{remark}$\Psi(x;\hbar)=\begin{pmatrix} \psi(x;\hbar)& \phi(x;\hbar)\\ \psi(x-\hbar;\hbar)&\phi(x-\hbar;\hbar)\end{pmatrix}$, i.e. $L(x;\hbar)=\begin{pmatrix} x& -1\\1&0\end{pmatrix}$, does not satisfy the parity property. Therefore, the choice of the constant $\lambda$ in the general setting of section \ref{blah} appears essential to obtain this property. However, the right choice may be hard to guess when $L_{1,2}(x;\hbar)$ has a series expansion in $\hbar$ involving even and odd powers.  
\end{remark}

\subsubsection{The leading order property}
In this section we prove the leading order condition of the topological type property. Following the idea of \cite{P2,Painleve}, we use the loop equations satisfied by the correlation functions:

\begin{proposition}[Theorem $2.9$ of \cite{Deter}]\label{Loop} Let us define the following functions (we denote by $L_n$ the set of variables $\{x_1,\dots,x_n\}$):
\bea \label{defP}
P_1(x)&=&\frac{1}{\hbar^2}\det D(x;\hbar)\cr
P_2(x;x_2)&=&
\frac{1}{\hbar} 
{\Tr}\left( \frac{D(x;\hbar)-D(x_2;\hbar)-(x-x_2)D'(x_2;\hbar)}{(x-x_2)^2}M(x_2;\hbar)\right)\cr
Q_{n+1}(x;L_n) \cr 
& & \hspace{-6.em} = \frac{1}{\hbar}\sum_{\sigma\in S_n} 
\frac{{\Tr}\left( D(x;\hbar)M(x_{\sigma(1)};\hbar)\dots 
M(x_{\sigma(n)};\hbar)\right)}{(x-x_{\sigma(1)})
(x_{\sigma(1)}-x_{\sigma(2)})\dots 
(x_{\sigma(n-1)}-x_{\sigma(n)})(x_{\sigma(n)}-x)} \cr
P_{n+1}(x;L_n)&=&
(-1)^n\left[Q_{n+1}(x;L_n)-\sum_{j=1}^n 
\frac{1}{x-x_j}\Res_{x'\to x_j} Q_{n+1}(x',L_n)\right]\cr
&&\eea
Then the correlation functions satisfy 
\beq \label{eq:first-loop-equation} P_1(x)=W_2(x,x;\hbar)+W_1(x;\hbar)^2, \eeq
and for $n\geq 1$:
\bea\label{ind1}
0&=&P_{n+1}(x;L_n)+W_{n+2}(x,x,L_n;\hbar)+2W_1(x;\hbar)W_{n+1}(x,L_n;\hbar)+\cr
&& \sum_{J \subset L_n, J\notin\{ \emptyset,L_n\}} W_{1+|J|}(x,J;\hbar)W_{1+n-|J|}(x,L_n\setminus J;\hbar) \cr
&&+ \sum_{j=1}^n \frac{d}{dx_j} \frac{ W_n(x,L_n\setminus x_j;\hbar)-W_n(L_n;\hbar)}{x-x_j}
\eea
Moreover $x\mapsto P_{n+1}(x;L_n)$ may only have singularities at the singularities of $D(x;\hbar)$.
\end{proposition}

We now combine the pole structure of the correlation functions $(W_n(x_1,\dots,x_n))_{n\geq 1}$ with the pole structure of the matrices $(D_k(x))_{k\geq 0}$ using theorems \ref{PoleD} and \ref{PoleStructureDiffSystem}. The second theorem and the loop equations (\eqref{eq:first-loop-equation} and \eqref{ind1}) shows that for $n\geq 0$, the functions $z\mapsto P_{n+1}(x(z);x_1,\dots,x_n)$ may only have poles at the branchpoints $z=\pm 1$. On the contrary, proposition \ref{Loop} combined with theorem \ref{PoleD} shows that $z\mapsto P_{n+1}(x(z);x_1,\dots,x_n)$ may only have singularity at $z\in\{0,\infty\}$. Consequently, the functions $z\mapsto P_{n+1}(x(z);x_1,\dots,x_n)$ are regular on $\bar{\mathbb{C}}$. Since the spectral curve is of genus $0$, this implies that they are constant.

\begin{theorem}\label{PP} The functions $x\mapsto P_{n+1}(x;L_n)$ do not depend on $x$. In other words, for all $n\geq 0$ we have:
\beqq P_{n+1}(x;x_1,\dots,x_n)=\td{R}_n(x_1,\dots,x_n)\eeqq
\end{theorem}  

The end of the proof of the leading order property is then similar to the one developed in \cite{P2,Painleve} for Painlev\'{e} equations that involves a clever induction. For completeness, we reproduce it in appendix \ref{AppendixA}. We end up with the following theorem:

\begin{theorem}\label{LeadingOrderDiffSystem} For any $n\geq 1$, the correlation function $W_n(x_1,\dots, x_n)$ defined from the system \eqref{DiffSystem} has a formal expansion in $\hbar$ starting at least at $\hbar^{n-2}$. 
\end{theorem}   

\subsubsection{Conclusion}
The results of the previous four sections prove that the differential system $\hbar \frac{d}{dx}\Psi(x;\hbar)=D(x;\hbar)\Psi(x;\hbar)$ arising from \eqref{DiffSystem} satisfy the topological type property. Consequently the correlation functions $(W_n(x_1,\dots,x_n))_{n\geq 1}$ defined by the determinantal formulas (definition \ref{alternative}) have a series expansion of the form:
\beqq W_n(x_1,\dots,x_n)=\sum_{g=0}^\infty W_n^{(n-2+2g)}(x_1,\dots,x_n)\hbar^{n-2+2g}\eeqq
where:
\beqq W_n^{(n-2+2g)}(x_1,\dots,x_n)dx_1\dots dx_n=\omega_n^{(g)}(x_1,\dots,x_n)\eeqq 
are the Eynard-Orantin differentials of the classical spectral curve \eqref{ClassicalCurve}.

\section{Comparing $M(x;\hbar)$ with the Dubrovin and Yang's conjecture\label{Comparision}}
From the previous section, we know that the determinantal formulas defined from the system \eqref{DiffSystem} identify to the Eynard-Orantin differentials that enumerates the Gromov-Witten invariants of $\mathbb{P}^1$. Thus, in order to prove Dubrovin and Yang conjecture, the last step is to compare the matrix $M(x;\hbar)$ defined in corollary \ref{ExpressionM} with the formulas proposed in \cite{Dubrovin}. The main difficulty here is that the conjecture of \cite{Dubrovin} only provides the series expansions of the matrices $\left(M_k(x)\right)_{k\geq 0}$ at $x\to \infty$. These large $x$ formal series have some interests in combinatorics since the coefficients usually happen to have some nice combinatorial interpretations. However, from the topological recursion point of view, the large $x$ expansions are very not well-suited in the formalism since they do not provide any information on the analytic structure that are required to take residues (in fact from the knowledge of the series, it is not obvious how to determine the radius of convergence and thus the location of the branchpoints).\\
In order to bypass this difficulty, we propose first to identify the matrix $M_0(x)$ with $\td{M}_0(x)$. Then, we shall prove that the conjectured matrix $\td{M}(x;\hbar)$ satisfy the equation $\delta_\hbar \td{M}(x;\hbar)=L(x;\hbar)\td{M}(x;\hbar)$ so that it must identify with $M(x;\hbar)$. We also mention that we could verify numerically the identification up to order $\hbar^6$ using results of section \eqref{SecSec}. 

\subsection{The $M_0(x)$ case \label{SectionM0}}
The first step is to compare our expression of $M_0(x)$ with the one conjectured in \cite{Dubrovin}. The expression proposed in \cite{Dubrovin} is:
\bea \left(\td{M}_0(x)\right)_{1,1}&=&1+\alpha_0(x)\cr
&=&1+\sum_{j=0}^\infty \frac{1}{4^jx^{2j+2}j!(j+1)!}\sum_{l=0}^j(-1)^l(2j+1-2l)^{2j+1}\binom{2j+1}{l}\cr
\left(\td{M}_0(x)\right)_{1,2}&=&-P_0(x)\cr
&=&-\sum_{j=0}^\infty\frac{1}{4^jx^{2j+1}(j!)^2}\sum_{l=0}^j (-1)^l(2j+1-2l)^{2j}\left(\binom{2j}{l}-\binom{2j}{l-1}\right)\cr
\left(\td{M}_0(x)\right)_{2,1}&=&P_0(x)\cr
\left(\td{M}_0(x)\right)_{2,2}&=&-\alpha_0(x)
\eea
Note in particular that the former expression satisfies $\alpha_0(x)=\frac{1}{2}(x-1)P_0(x)$. This is compatible with the structure of $M_0(x)$ given in corollary \eqref{ExpressionM}. Therefore, the only remaining point is to prove that the asymptotic expansion at $x\to \infty$ of $\frac{1}{\sqrt{x^2-4}}$ equals $P_0(x)$. We have:
\beq \frac{1}{\sqrt{x^2-4}}=\sum_{k=0}^\infty\frac{(2k)!}{(k!)^2x^{2k+1}}\eeq
Eventually, the final step is to observe that the following identity holds (easily proven by induction on $j\geq 0$):
\beq \forall\, j\geq 0\,:\, \sum_{l=0}^j \frac{(-1)^l(2j+1-2l)^{2j+1}(2j)!}{l!(2j-l+1)!}=4^j\eeq
so that $P_0(x)$ corresponds to the expansion of $\frac{1}{\sqrt{x^2-4}}$ at $x\to \infty$. Therefore, our matrix $M_0(x)$ matches the one conjectured in \cite{Dubrovin}.

\subsection{The $(M_k(x))_{k\geq 1}$ cases}
In order to compare our matrix with the one proposed in \cite{Dubrovin}, we rewrite the formulas of conjecture \eqref{ConjectureM}. We first observe:
\beq \binom{2i}{l}-\binom{2i}{l-1}=\frac{(2i)!(2i-2l+1)!}{(2i-l+1)! (l!)}\,\,,\,\, \forall\, (i,l)\in\mathbb{N}\times\mathbb{N}^*\eeq
with the convention that $\binom{n}{-1}=0$ for all $n\geq 0$. Thus we get:

\bea \label{Rewriting}\alpha(x;\hbar)&=&\sum_{m=0}^\infty\sum_{s=0}^\infty\sum_{l=0}^{s}\frac{(-1)^l(2s+1-2l)^{2s+2m+1}}{4^{s+m}(s!)(s+1)!}\binom{2s+1}{l}\frac{\hbar^{2m}}{x^{2s+2m+2}}\cr
P(x;\hbar)&=&\sum_{m=0}^\infty\sum_{s=0}^\infty\sum_{l=0}^{s}\frac{(-1)^l(2s+1-2l)^{2s+2m+1}}{4^{s+m}(s!)^2(2s+1)}\binom{2s+1}{l} \frac{\hbar^{2m}}{x^{2s+2m+1}}\cr
Q(x;\hbar)&=&\frac{1}{2}\sum_{m=0}^\infty\sum_{s=0}^\infty\sum_{l=0}^{s}\frac{(-1)^l(2s+1-2l)^{2s+2m+1}}{4^{s+m}(s!)^2}\binom{2s+1}{l}\frac{\hbar^{2m+1}}{x^{2s+2m+2}}\cr &&
\eea
We now use the fact that:
\beq \frac{d^k}{dx^k}\frac{1}{x^m}=\frac{(-1)^k(m+k-1)!}{(m-1)! \,x^{m+k}} \,\,\,,\,\, \forall \, (m,k)\in \mathbb{N}^*\times \mathbb{N}\eeq
to compute $\alpha(x+\hbar;\hbar)$, $P(x+\hbar;\hbar)$ and $Q(x+\hbar;\hbar)$: 
\footnotesize{\bea \label{XPLUSH}
\alpha(x+\hbar;\hbar)&=&\sum_{p=0}^\infty\sum_{s=0}^\infty\sum_{l=0}^s\sum_{m=0}^p\frac{(-1)^l(2s+1-2l)^{2s+1+2p-2m}}{4^{s+p-m}(s!)(s+1)!}\binom{2s+1}{l}\binom{2s+1+2p}{2m}\frac{\hbar^{2p}}{x^{2s+2p+2}}\cr
&&-\sum_{p=0}^\infty\sum_{s=0}^\infty\sum_{l=0}^s\sum_{m=0}^{p}\frac{(-1)^l(2s+1-2l)^{2s+1+2p-2m}}{4^{s+p-m}(s!)(s+1)!}\binom{2s+1}{l}\binom{2s+2+2p}{2m+1}\frac{\hbar^{2p+1}}{x^{2s+2p+3}}\cr
P(x+\hbar;\hbar)&=&\sum_{p=0}^\infty\sum_{s=0}^\infty\sum_{l=0}^s\sum_{m=0}^p\frac{(-1)^l(2s+1-2l)^{2s+1+2p-2m}}{4^{s+p-m}(s!)^2(2s+1)}\binom{2s+1}{l}\binom{2s+2p}{2m}\frac{\hbar^{2p}}{x^{2s+2p+1}}\cr
&&-\sum_{p=0}^\infty\sum_{s=0}^\infty\sum_{l=0}^s\sum_{m=0}^p\frac{(-1)^l(2s+1-2l)^{2s+1+2p-2m}}{4^{s+p-m}(s!)^2(2s+1)}\binom{2s+1}{l}\binom{2s+1+2p}{2m+1}\frac{\hbar^{2p+1}}{x^{2s+2p+2}}\cr
Q(x+\hbar;\hbar)&=&\frac{1}{2}\sum_{p=0}^\infty\sum_{s=0}^\infty\sum_{l=0}^s\sum_{m=0}^p\frac{(-1)^l(2s+1-2l)^{2s+1+2p-2m}}{4^{s+p-m}(s!)^2}\binom{2s+1}{l}\binom{2s+2p+1}{2m}\frac{\hbar^{2p+1}}{x^{2s+2p+2}}\cr
&&-\frac{1}{2}\sum_{p=0}^\infty\sum_{s=0}^\infty\sum_{l=0}^s\sum_{m=0}^{p-1}\frac{(-1)^l(2s+1-2l)^{2s-1+2p-2m}}{4^{s+p-m-1}(s!)^2}\binom{2s+1}{l}\binom{2s+2p}{2m+1}\frac{\hbar^{2p}}{x^{2s+2p+1}}\cr
&&
\eea
}\normalsize{}
The strategy is then to verify that the matrix conjectured in \cite{Dubrovin} satisfy the identities:
\beq \label{Conditions}M(x+\hbar;\hbar)L(x;\hbar)=L(x;\hbar)M(x;\hbar) \text{ and } \det M(x;\hbar)=0 \text{ and }\Tr M(x;\hbar)=1\eeq
We first note that $\td{M}(x:\hbar)$ defined in \eqref{ConjectureM} obviously satisfies the trace condition. The condition on the determinant can be deduced from the first condition. Indeed, if $\td{M}(x;\hbar)$ satisfies the identity $\td{M}(x+\hbar;\hbar)=L(x;\hbar)\td{M}(x;\hbar)L^{-1}(x;\hbar)$, then the determinant $d(x;\hbar)\overset{\text{def}}{=}\det \td{M}(x;\hbar)$ satisfies the equation:
\bea \delta_\hbar d(x;\hbar)&=&d(x+\hbar;\hbar)=\det(M(x+\hbar;\hbar))=\det\left(L(x;\hbar)M(x;\hbar)L^{-1}(x;\hbar)\right)\cr
&=&d(x;\hbar)\eea
Projecting the last identity at order $\hbar^k$ with $k\geq 1$ gives:
\beq\sum_{j=1}^k \frac{d^j}{dx^j} d_{k-j}(x)=0\eeq
Thus, we obtain:
\bea \label{rtt} d_0'(x)&=&0\cr
d_1'(x)&=&-d_0''(x)\cr
d_{k}'(x)&=&-\sum_{j=1}^{k}\frac{d^j}{dx^j}d_{k-j}(x) \,\,,\,\, \forall \, k\geq 1
\eea
From section \ref{SectionM0}, we know that the matrix of \cite{Dubrovin} is the same as \eqref{M0}. In particular, we have $d_0(x)=0$. This implies from \eqref{rtt} that $d_1'(x)=0$. Then by a straightforward induction, we obtain $d_k'(x)=0$ for all $k\geq 1$. Eventually, from definition \eqref{ConjectureM2}, we get that:
\beq \forall \, k\geq 1\,\,:\,\, \td{M}_k(x)\underset{x\to \infty}{\to} 0\eeq
Combining the last identity with the fact that $d_k'(x)=0$ for all $k\geq 1$ implies that:
\beq \forall \, k\geq 0\,:\, d_k(x)=0 \,\,\Leftrightarrow \,\,\, \det \td{M}(x;\hbar)=0\eeq
Therefore, the only remaining issue is to prove that the matrix $\td{M}(x;\hbar)$ proposed in equation \eqref{ConjectureM} satisfies $\td{M}(x+\hbar;\hbar)L(x;\hbar)=L(x;\hbar)\td{M}(x;\hbar)$. Since the trace is fixed to $1$, this is equivalent to verify that the relation holds for the entries $(1,1)$, $(1,2)$ and $(2,1)$. Using the specific form of the matrix \eqref{ConjectureM}, this is equivalent to prove that:
\bea \label{SystemeReduit}\alpha(x+\hbar)_{2p+1}&=&-xQ(x)_{2p}+\frac{1}{2}P(x)_{2p}\cr
\alpha(x+\hbar)_{2p}&=&-\frac{1}{2}Q_{2p-1}(x)+xP_{2p}(x)-\alpha(x)_{2p}\cr
P(x)_{2p}&=&Q(x+\hbar)_{2p}+P(x+\hbar)_{2p}\cr
-Q(x)_{2p+1}&=&Q(x+\hbar)_{2p+1}+P(x+\hbar)_{2p+1}\cr
\alpha(x)_{2p}&=&xQ(x+\hbar)_{2p}+xP(x+\hbar)_{2p}+\frac{1}{2}Q(x+\hbar)_{2p-1}\cr
&&+\frac{1}{2}P(x+\hbar)_{2p-1}-\alpha(x+\hbar)_{2p}\cr
\alpha(x)_{2p+1}&=&xQ(x+\hbar)_{2p+1}+xP(x+\hbar)_{2p+1}+\frac{1}{2}Q(x+\hbar)_{2p}\cr
&&+\frac{1}{2}P(x+\hbar)_{2p}-\alpha(x+\hbar)_{2p+1}
\eea
Using \eqref{Rewriting} and \eqref{XPLUSH}, we get:
\begin{itemize}\item Equation $\alpha(x+\hbar)_{2p+1}=-xQ(x)_{2p}+\frac{1}{2}P(x)_{2p}$ is equivalent to:
\bea&&\sum_{l=0}^{s-1}\sum_{m=0}^p (-1)^l 4^{m+1}(2s-1-2l)^{2s+2p-2m-1}\binom{2s-1}{l}\binom{2s+2p}{2m+1}=\cr
&&\frac{1}{2s+1}\sum_{l=0}^s(-1)^l(2s+1-2l)^{2s+2p+1}\binom{2s+1}{l}\cr
&&
\eea
\item Equation $\alpha(x+\hbar)_{2p}=-\frac{1}{2}Q_{2p-1}(x)+xP_{2p}(x)-\alpha(x)_{2p}$ is equivalent to:
\bea 
&&\sum_{l=0}^s\sum_{m=0}^p(-1)^l4^m(2s+1-2l)^{2s+1+2p-2m}\binom{2s+1}{l}\binom{2s+1+2p}{2m}=\cr
&&+\sum_{l=0}^{s+1}\frac{(-1)^l(2s+3-2l)^{2s+2p+1}}{4(s+1)}\left(\frac{(2s+3-2l)^2}{2s+3}-1\right)\binom{2s+3}{l}\cr
&&-\sum_{l=0}^s(-1)^l(2s+1-2l)^{2s+2p+1}\binom{2s+1}{l}\cr
&&
\eea
\item Equation $P(x)_{2p}=Q(x+\hbar)_{2p}+P(x+\hbar)_{2p}$ is equivalent to:
\bea
&&\sum_{l=0}^s(-1)^l(2s+1-2l)^{2s+2p+1}\binom{2s+1}{l}=\cr
&&-2\sum_{l=0}^s\sum_{m=0}^{p-1}(-1)^l 4^m(2s+1-2l)^{2s-1+2p-2m}(2s+1)\binom{2s+1}{l}\binom{2s+2p}{2m+1}\cr
&&\sum_{l=0}^s\sum_{m=0}^{p}(-1)^l 4^m(2s+1-2l)^{2s+1+2p-2m}\binom{2s+1}{l}\binom{2s+2p}{2m}\cr
&&
\eea
\item Equation $-Q(x)_{2p+1}=Q(x+\hbar)_{2p+1}+P(x+\hbar)_{2p+1}$ is equivalent to:
\bea
&&\sum_{l=0}^s(-1)^l(2s+1-2l)^{2s+2p+1}\binom{2s+1}{l}=\cr
&&-\sum_{l=0}^s\sum_{m=0}^{p}(-1)^l 4^m(2s+1-2l)^{2s+1+2p-2m}\binom{2s+1}{l}\binom{2s+2p+1}{2m}\cr
&&+2\sum_{l=0}^s\sum_{m=0}^{p}\frac{(-1)^l 4^m(2s+1-2l)^{2s-1+2p-2m}}{2s+1}\binom{2s+1}{l}\binom{2s+2p+1}{2m+1}\cr
&&
\eea
\item Equation $\alpha(x)_{2p}=xQ(x+\hbar)_{2p}+xP(x+\hbar)_{2p}+\frac{1}{2}Q(x+\hbar)_{2p-1}+\frac{1}{2}P(x+\hbar)_{2p-1}-\alpha(x+\hbar)_{2p}$ is equivalent to:
\bea
&&4s\sum_{l=0}^{s-1}(-1)^l(2s-1-2l)^{2s+2p-1}\binom{2s-1}{l}=\cr
&&-2\sum_{l=0}^s\sum_{m=0}^{p-1}(-1)^l 4^m(2s+1-2l)^{2s-1+2p-2m}\binom{2s+1}{l}\binom{2s+2p}{2m+1}\cr
&&+\sum_{l=0}^s\sum_{m=0}^p \frac{(-1)^l4^m(2s+1-2l)^{2s+1+2p-2m}}{2s+1}\binom{2s+1}{l}\binom{2s+2p}{2m}\cr
&&+\sum_{l=0}^s\sum_{m=0}^{p-1}(-1)^l4^m(2s+1-2l)^{2s-1+2p+2m}\binom{2s+1}{l}\binom{2s+2p-1}{2m}\cr
&&-2\sum_{l=0}^s\sum_{m=0}^{p-1}\frac{(-1)^l4^m(2s+1-2l)^{2s-1+2p+2m}}{2s+1}\binom{2s+1}{l}\binom{2s+2p-1}{2m+1}\cr
&&-4s\sum_{l=0}^{s-1}\sum_{m=0}^p(-1)^l 4^m(2s-1-2l)^{2s+2p-2m-1}\binom{2s-1}{l}\binom{2s+2p-1}{2m}\cr
&&
\eea
\item Equation $\alpha(x)_{2p+1}=xQ(x+\hbar)_{2p+1}+xP(x+\hbar)_{2p+1}+\frac{1}{2}Q(x+\hbar)_{2p}+\frac{1}{2}P(x+\hbar)_{2p}-\alpha(x+\hbar)_{2p+1}$ is equivalent to:
\bea
&&0=\frac{1}{2}\sum_{l=0}^s\sum_{m=0}^p(-1)^l4^m(2s+1-2l)^{2s+1+2p-2m}\binom{2s+1}{l}\binom{2s+2p+1}{2m}\cr
&&-\sum_{l=0}^s\sum_{m=0}^p\frac{(-1)^l4^m(2s+1-2l)^{2s+1+2p-2m}}{2s+1}\binom{2s+1}{l}\binom{2s+1+2p}{2m+1}\cr
&&-\sum_{l=0}^s\sum_{m=0}^{p-1}(-1)^l4^m(2s+1-2l)^{2s-1+2p-2m}\binom{2s+1}{l}\binom{2s+2p}{2m+1}\cr
&&+\frac{1}{2}\sum_{l=0}^s\sum_{m=0}^p \frac{(-1)^l4^m(2s+1-2l)^{2s+1+2p-2m}}{2s+1}\binom{2s+1}{l}\binom{2s+2p}{2m}\cr
&&+4s\sum_{l=0}^{s-1}\sum_{m=0}^p(-1)^l4^m(2s-1-2l)^{2s-1+2p-2m}\binom{2s-1}{l}\binom{2s+2p}{2m+1}\cr
&&
\eea
\end{itemize}

All six identities can easily been proved using standard relations on binomial coefficients and hypergeometric functions \cite{Gould}. Therefore, we conclude that conjecture \eqref{ConjectureM} adapted from \cite{Dubrovin} is verified. We could also verify with formal numeric computations that the correlation functions identify with the corresponding Eynard-Orantin differentials up to $n+2g-2\leq 4$. In particualr, the first Eynard-Orantin differentials are:
\small{\beaa \omega_2^{(0)}(z_1,z_2)&=&\frac{dz_1dz_2}{(z_1-z_2)^2}\cr
\omega_3^{(0)}(z_1,z_2,z_3)&=&\frac{dz_1dz_2dz_3}{2(z_1-1)^2(z_2-1)^2(z_3-1)^2}+\frac{dz_1dz_2dz_3}{2(z_1+1)^2(z_2+1)^2(z_3+1)^2}\cr
\omega_1^{(1)}(z)&=&-\frac{1}{48}\left(\frac{dz}{(z-1)^2}+\frac{dz}{(z+1)^2}\right)+\frac{1}{16}\left(\frac{dz}{(z-1)^3}-\frac{dz}{(z+1)^3}\right)\cr
&&+\frac{1}{16}\left(\frac{dz}{(z-1)^4}+\frac{dz}{(z+1)^4}\right)\cr
\omega_1^{(2)}(z)&=&\frac {{z}^{2}({z}^{2}+1)(7{z}^{12}-52{z}^{10}+7985{z}^{8}+34520{z}^{6}+7985{z}^{4}-52{z}^{2}+7) }{ 960(z-1)^{10}  (z+1 )^{10}}dz
\eeaa}\normalsize{}


\section{Conclusion and outlooks}
The purpose of this article was to prove on the simple example of the quantum curve arising in the enumeration of Gromov-Witten invariants of $\mathbb{P}^1$ that the determinantal formulas and the topological type property may be used in the context of $\hbar$-difference systems rather than $\hbar$-differential systems. However, as presented in this paper, it seems that the construction might be adapted for more general situations. In particular, it raises the following questions:
\begin{itemize}\item Formulas presented in section \ref{DL} are valid for any $2\times 2$ $\hbar$-difference systems. Therefore it seems that the topological type property might be proven for a wide class of $2\times 2$ matrices $L(x;\hbar)$ that are rational in $x$. This would be the generalization of \cite{LoopLie} for $2\times 2$ $\hbar$-difference systems.
\item As mentioned in the article, the present situation may be seen as a specific connection on the Lie group $G=SL_2(\mathbb{C})$ described by the rational matrix $L(x;\hbar)$. From the results of \cite{LoopLie2}, it seems reasonable to believe that the present strategy might be extended to other Lie groups and other connections.
\item At the level of operators, $\delta_\hbar$ is the exponentiated version of the operator $\hbar\partial_x$, but at the level of matrices (local representations), the relation is only correct for the leading order ($L_0(x)=\text{exp}(D_0(x))$). For higher orders, the situation seems more involved and would deserve algebraic investigations.
\item There are many examples of hyperbolic classical spectral curves arising in the context of enumerative geometry (like for example Gromov-Witten invariants on toric Calabi-Yau manifolds) \cite{Panda,Dessin2,Quantum,ListSpecCurve,Eguchi,DZnew,OP,GP,DubrovinSolo} for which the present situation might be of interest. In particular, it would be interesting to build the associated $\hbar$-difference systems associated and check if the topological type property holds.
\item In this article, we only considered formal WKB series expansion in $\hbar$. Though this perspective is sufficient for enumerative geometry (where formal series are perfectly adapted), the issue of the convergence of the series and the connection with so-called exact WKB expansions \cite{IwakiExactWKB,IwakiExactWKB2} would deserve investigations.
\item From the topological recursion point of view, we observe that the spectral curve arising in the $\mathbb{P}^1$ case:
\beqq x(z)=z+\frac{1}{z} \,\,\,,\,\, y(z)=\ln z\eeqq
is obtained from the curve $\left(\td{y}-\frac{\td{x}}{2}\right)=\frac{\td{x}^2}{4}-1$ with the change of variables $(\td{x},\td{y})=\left(x,e^{y}\right)$ whose quantization is a Schr\"{o}dinger differential equation. It would be interesting to see if any relation between the two sets of Eynard-Orantin differentials exists and if the previous change of variables may be used for other spectral curves.  
\end{itemize}

\section*{Acknowledgment}
The author would like to thank Taro Kimura for fruitful discussion. O.~Marchal would like to thank Universit\'e Lyon $1$, Universit\'e Jean Monnet and Institut Camille Jordan for material support. O.~Marchal would also like to thank Pauline Chappuis for moral support during the preparation of this article. O.~Marchal is supported by the LABEX MILYON (ANR-10-LABX-0070) of Universit\'e de Lyon, within the program ``Investissements d'Avenir'' (ANR-11-IDEX-0007) operated by the French National Research Agency (ANR).

\renewcommand{\thesection}{\Roman{section})}
\renewcommand{\theequation}{\thesection.\arabic{equation}}
\begin{appendices}

\section{\label{AppendixA}Appendix: induction for the leading order property}
In this appendix, we present the direct adaptation of the induction proof presented in \cite{P2} and \cite{Painleve} that can be carried out from theorem \ref{PP}.

Let us define the following statement:
\beq \mathcal{P}_k \,:\,  
W_j(x_1,\dots,x_j) \text{ is at least of order } \hbar^{k-2} \quad \text{for $j \ge k$}. \eeq
The statement is obviously true for $k=1$ and $k=2$ from the definitions of the correlation functions. Let us assume that the statement $\mathcal{P}_i$ is true for all $i\leq n$. Now we look at the loop equations \eqref{ind1} that we recall for $n\geq 1$:
\bea
0&=&P_{n+1}(x;L_n)+W_{n+2}(x,x,L_n)+2W_1(x)W_{n+1}(x,L_n)+\cr
&& \sum_{J \subset L_n, J\notin\{ \emptyset,L_n\}} W_{1+|J|}(x,J)W_{1+n-|J|}(x,L_n\setminus J) \cr
&&+ \sum_{j=1}^n \frac{d}{dx_j} \frac{ W_n(x,L_n\setminus x_j)-W_n(L_n)}{x-x_j}
\eea
By the induction assumption, we have that the last two terms are at least of order $\hbar^{n-2}$. Indeed, in the sum we have terms of order $\hbar^{1+|J|-2+1+n-|J|-2}=\hbar^{n-2}$. Moreover, we also have from the same assumption that $W_{n+2}(x,x,L_n)$ is also of order at least $\hbar^{n-2}$ (since $n+2\geq n$). Therefore, $W_{n+1}(x,L_n)$ is at least of order $\hbar^{n-2}$, and by considering the coefficients of the $\hbar^{n-3}$ we have:
\beq \label{eqsys} 0=P_{n+1}^{(n-3)}(x;L_n)+2W_1^{(-1)}(x)W_{n+1}^{(n-2)}(x,L_n) \eeq
If we assume that $W_{n+1}^{(n-2)}(x,L_n)\neq 0$ (hence $P_{n+1}^{(n-3)}(x;L_n)\neq 0$), then we have
\beq W_{n+1}^{(n-2)}(x,L_n)=\frac{P_{n+1}^{(n-3)}(x;L_{n})}{2W_1^{(-1)}(x)}. \eeq
Since $W_1^{(-1)}(x)$ is the spectral curve of the system by definition and $P_{n+1}(x;L_n)$ does not depend on $x$ (theorem \ref{PP}), we obtain:
\beq W_{n+1}^{(n-2)}(x,L_{n})=\frac{P_{n+1}^{(n-3)}(L_{n})}{2\ln\left(\frac{x}{2}+\frac{1}{2}\sqrt{x^2-4}\right)}\overset{x\to \infty}{=}\frac{P_{n+1}^{(n-3)}(L_{n})}{x}+O\left(\frac{1}{x^3}\right)\eeq
This would imply that $x\mapsto W_{n+1}^{(n-2)}(x,L_{n})$ has a non-zero residue at $x\to \infty$ in contradiction with its pole structure and its asymptotic at $x\to \infty$ given by theorem \ref{PoleStructureDiffSystem}. Consequently, we must have $W_{n+1}^{(n-2)}(x,L_{n})=0$. This proves that $W_{n+1}(x,L_n)$ is at least of order $\hbar^{n-1}$. We now need to prove the same statement for higher correlation functions. Let us prove it by a second induction by defining
\beq \td{\mathcal{P}}_i \,:\, W_i(x_1,\dots,x_i) \text{ is of order at least } \hbar^{n-1}. \eeq
We prove $\td{\mathcal{P}}_i$ for all $i\geq n+1$ by induction. We just proved it for $i=n+1$ so initialization is done. Let us assume that $\td{\mathcal{P}}_j$ is true for all $j$ satisfying $n+1\leq j\leq i_0$. We look at the loop equation:
\bea \label{ind2} 0&=&P_{i_0+1}(x;L_{i_0})+W_{i_0+2}(x,x,L_{i_0})+2W_1(x)W_{i_0+1}(x,L_{i_0})\cr
&&+ \sum_{J \subset L_{i_0}, J\notin\{ \emptyset,L_{i_0}\}} W_{1+|J|}(x,J)W_{1+i_0-|J|}(x,L_{i_0}\setminus J)\cr
&&+\sum_{j=1}^{i_0} \frac{d}{dx_j} \frac{ W_{i_0}(x,L_{i_0}\setminus x_j)-W_{i_{0}}(L_{i_0})}{x-x_j}. 
\eea
By assumption on $\td{\mathcal{P}}_{i_0}$, the last sum with the derivatives contains terms of order at least $\hbar^{n-1}$. In the sum involving the subsets of $L_{i_0}$, it is straightforward to see that the terms are all of order at least $\hbar^{n-1}$. Indeed, as soon as one of the index is greater than $n+1$, the assumption $\td{\mathcal{P}}_i$ for $n+1\leq i\leq i_0$ tells us that this term is already at order at least $\hbar^{n-1}$. Since the second factor of the product is at least of order $\hbar^0$, it does not decrease the order. Now, if both factors have indexes strictly lower than $n+1$, then the assumption of $\mathcal{P}_j$  for all $j\leq n$ tell us that the order of the product is at least of $\hbar^{|J|+1-2+1+i_0-|J|-2}=\hbar^{i_0-2}$ which is greater than $n-1$ since $i_0\geq n+1$. Additionally, by induction on $\mathcal{P}_n$ we know that $W_{i_0+1}(x,L_{i_0})$ is at least of order $\hbar^{n-2}$ as well as $W_{i_0+2}(x,x,L_{i_0})$. Consequently, looking at order $\hbar^{n-3}$ in \eqref{ind2} gives
\beq \label{eqsys2} 0=P_{i_0+1}^{(n-3)}(x;L_{i_0})+2W_1^{(-1)}(x)W_{i_0+1}^{(n-2)}(x,L_{i_0}) \eeq
We can apply a similar reasoning as the one developed for \eqref{eqsys}. If we assume $W_{i_0+1}^{(n-2)}(x,L_{i_0})\neq 0$ (hence $P_{i_0+1}^{(n-3)}(x;L_{i_0})\neq 0$), then we have:
\beq W_{i_0+1}^{(n-2)}(x,L_{i_0})=\frac{P_{i_0+1}^{(n-3)}(L_{i_0})}{2W_1^{(-1)}(x)} \eeq
Again this would imply that $x\mapsto W_{i_0+1}^{(n-2)}(x,L_{i_0})$ has a non-zero residue at $x\to \infty$ in contradiction with its pole structure and its asymptotic at $x\to \infty$ given by theorem \ref{PoleStructureDiffSystem}. Consequently, we must have $W_{i_0+1}^{(n-2)}(x,L_{i_0})=0$. In particular, it means that $W_{i_0+1}(x,L_{i_0})$ (which by assumption of $\mathcal{P}_n$ was already known to be at least of order $\hbar^{n-2}$) is at least of order $\hbar^{n-1}$, making the induction on $\td{\mathcal{P}}_{i_0}$. Hence, by induction, we have proved that, $\forall\, i \geq n+1$, $\td{\mathcal{P}}_{i}$ holds which proves that $\mathcal{P}_{n+1}$ is true. Then, by induction, we have just proved that $\mathcal{P}_{n}$ holds for $n \geq 1$. In other words, we have proved the leading order condition of the topological type property for the system defined by \eqref{DiffSystem}.

\end{appendices}
\end{document}